\documentclass[aps,prl,superscriptaddress,twocolumn,showpacs]{revtex4-1}

\usepackage[utf8]{inputenc}
\usepackage[T1]{fontenc}
\usepackage[german,english]{babel}

\usepackage{amsmath}
\usepackage{mathtools}
\usepackage{amssymb}
\usepackage{amsthm}
\usepackage{microtype} 
\usepackage{quoting} 
\quotingsetup{font=small}

\usepackage[pdftex]{graphicx}
\usepackage{fancyhdr}
\usepackage{bm}
\usepackage{comment}
\usepackage{booktabs}
\usepackage{eurosym} 
\usepackage{braket}
\usepackage{bbold} 
\usepackage{emptypage}
\usepackage{epstopdf}
\usepackage{verbatim}
\usepackage{enumitem}
\usepackage{float} 
\usepackage{chemformula}

\definecolor{islamicgreen}{rgb}{0.0, 0.56, 0.0}
\newcommand{\review}[1]{{
\textcolor{blue}{#1}}}

\newcommand{\beq}{\begin{equation}}
\newcommand{\eeq}{\end{equation}}
\newcommand {\nn} \nonumber

\renewcommand{\ket}[1]{\left|#1\right\rangle}

\DeclarePairedDelimiter{\abs}{\lvert}{\rvert} 

\DeclareMathOperator{\Max}{Max}

\renewcommand{\sim}{\thicksim}

{\left\lbrace\begin{array}{@{}l@{}}}%
{\end{array}\right.}

\usepackage[bookmarks=true,colorlinks,linkcolor=blue,urlcolor=blue,citecolor=blue]{hyperref} 

\begin{document}

\title{
{Quasilocalized} dynamics from confinement of quantum excitations
}
\author{Alessio Lerose}
\thanks{These authors equally contributed to this work.}
\affiliation{SISSA -- International School for Advanced Studies, via Bonomea 265, 34136 Trieste, Italy.}
\affiliation{INFN, Sezione di Trieste, via Bonomea 265, 34136 Trieste, Italy.}
\author{Federica M.~Surace}
\thanks{These authors equally contributed to this work.}
\affiliation{SISSA -- International School for Advanced Studies, via Bonomea 265, 34136 Trieste, Italy.}
\affiliation{ICTP -- International Center for Theoretical Physics, Strada Costiera 11, 34151 Trieste, Italy.}
\author{Paolo P. Mazza}
\affiliation{SISSA -- International School for Advanced Studies, via Bonomea 265, 34136 Trieste, Italy.}
\affiliation{INFN, Sezione di Trieste, via Bonomea 265, 34136 Trieste, Italy.}
\author{Gabriele Perfetto}
\affiliation{SISSA -- International School for Advanced Studies, via Bonomea 265, 34136 Trieste, Italy.}
\affiliation{INFN, Sezione di Trieste, via Bonomea 265, 34136 Trieste, Italy.}
\author{Mario Collura}
\affiliation{SISSA -- International School for Advanced Studies, via Bonomea 265, 34136 Trieste, Italy.}
\author{Andrea Gambassi}
\affiliation{SISSA -- International School for Advanced Studies, via Bonomea 265, 34136 Trieste, Italy.}
\affiliation{INFN, Sezione di Trieste, via Bonomea 265, 34136 Trieste, Italy.}
\date{\today} 

\begin{abstract}
Confinement of excitations induces {quasilocalized} dynamics in
   {disorder-free} isolated quantum {many-body} 
   systems in one spatial dimension.
   This occurrence is
    signalled by severe suppression of quantum correlation spreading and of entanglement growth, long-time persistence of spatial inhomogeneities, and long-lived coherent oscillations 
    of local observables. 
    In this work, we {present a unified understanding} of these dramatic effects.
     The slow dynamical behavior is shown to be related to the Schwinger effect in quantum electrodynamics.
    We demonstrate that it is quantitatively captured for long time scales by effective Hamiltonians exhibiting Stark localization of excitations 
    and weak growth of the entanglement entropy for arbitrary coupling strength.
    This analysis explains the phenomenology of real-time string dynamics investigated in a number of lattice gauge theories, 
    as well as the anomalous dynamics observed in quantum Ising chains after quenches. 
    Our findings establish confinement as a robust mechanism for hindering 
    the approach to equilibrium in translationally-invariant quantum statistical systems with local interactions.
\end{abstract}

\maketitle

\emph{Introduction.}— 
Elementary particles such as quarks experience spatial confinement into composite particles, due to forces acting at arbitrary distances mediated by gauge fields \cite{Weinbergbook2}.
 An analogous effect is also present in condensed-matter systems. 
{In one spatial dimension,} confinement typically arises in the ordered phases of systems with a spontaneously broken discrete symmetry: their elementary particle/antiparticle excitations consist of kink/antikink configurations which locally connect different degenerate ground states (vacua).
Upon breaking the symmetry via external fields,  the various vacua acquire  different energy densities.
As a result, separating a kink-antikink pair requires a configurational energy cost which grows proportionally to their distance \cite{McCoyWuConfinement,80_SpinonConfinement,DelfinoMussardoSimonetti,DelfinoMussardo,Affleck_SolitonConfinementTheory,Affleck_SolitonSpinPeierlsNumerical,Augier_SolitonBoundStatesRaman,ShankarConfinement,RutkevichFalseVacuum,DelfinoDecayIFF,FonsecaZamMesonSpectrumContinuum,RutkevichMesonSpectrumContinuum,ShollwockConfinement,RutkevichMesonSpectrumLattice,RutkevichConfinementXXZ,SuzukiConfinementSpin1,WangSpinonConfinement,BeraEsslerSpinonConfinement}.

Several recent numerical studies of one-dimensional lattice gauge theories 
and quantum spin chains 
have found that confinement may give rise to anomalous real-time dynamics \cite{WilczekCoherentOscillationsQCD,HebenstreitRealTimeStringBreakingLong,KCTC,BuyensRealTimeSchwingerTensorNetwork,BanulsWeakThermalization,LinMotrunichOscillations,MazzaTransport,PichlerMontangeroQLMTensorNetwork,KunoQSimul1dHiggs2,HebenstreitRealTimeStringBreaking,SalaVariationalStudy,ParkGlassyQuark,NotarnicolaZ3Rydberg,SuraceRydberg,GorshkovConfinement,LeroseDWLR,Ercolessi2019, Tagliacozzo2019,PaiPretkoFractonsLGT}
and spectral properties \cite{RobinsonNonthermalStatesShort,RobinsonNonthermalStatesLong,CuberoScars}  at finite energy density above the ground state, in contrast with the 
generically expected thermalization \cite{ShankarStatisticalNEQ,Rigol2008,Eisert2015,HuseLebowitz,MoriReview,DeutschETH,SrednickiETH,DAlessio_ETHReview}.
The signatures of these phenomena 
include
extraordinary long-lived coherent oscillations of local observables \cite{WilczekCoherentOscillationsQCD,HebenstreitRealTimeStringBreakingLong,KCTC,BuyensRealTimeSchwingerTensorNetwork,BanulsWeakThermalization,LinMotrunichOscillations,GorshkovConfinement,Delfino_2014}, suppression of the light-cone spreading of quantum correlations \cite{KCTC,GorshkovConfinement} and of the entanglement growth \cite{PichlerMontangeroQLMTensorNetwork,KCTC,GorshkovConfinement}, and persistent  
inhomogeneities \cite{MazzaTransport,PichlerMontangeroQLMTensorNetwork,KunoQSimul1dHiggs2,HebenstreitRealTimeStringBreaking,SalaVariationalStudy,ParkGlassyQuark,NotarnicolaZ3Rydberg,SuraceRydberg,LeroseDWLR}. 
While these observations suggest that confinement is related to a  suppression of thermalization, the nature of this connection has not yet been clarified.

In this work we investigate  the relationship between the aforementioned dynamical effects of confinement  and prototypical aspects of the localization of interacting particles \cite{AndersonLocalization,BAA,HuseOganesyanMBL,HusePhenomenologyMBL,ProsenMBL,BardarsonPollmanMoore,VasseurMooreReview,Serbyn2013universal,SPACriterionMBL,DeRoeckHuveneersScenario,DeRoeck1,Schiulaz1,Cirac:QuasiMBLWithoutDisorder,CarleoGlassy,BolsDeRoeckBoseHubbard,GarrahanMBLGlass,GarrahanQuasiMBLEast,AbaninProsen:SlowDynamics,QuasiMBLFrystration,RefaelStarkMBL,SchultzStarkMBL}.
We demonstrate that confinement  causes quasilocalized dynamics of states with dilute excitations. 
In fact, the route towards thermalization involves the decay of these 
states 
into entropically favored many-particle states: 
the energy stored in confining strings has to be converted into mass via the creation of new pairs of excitations from the vacuum.
We show that these processes can become dramatically slow, 
in close analogy with the Schwinger \mbox{effect}, i.e., with the suppressed decay of false vacua in quantum electrodynamics \cite{SchwingerMechanism}.
In this regime, 
fast spatial propagation of excitations is prevented by their {Stark} localization \cite{WannierStarkLocalization} in the mutual confining potentials.
\begin{figure} 
\centering
\includegraphics[width=0.48\textwidth]{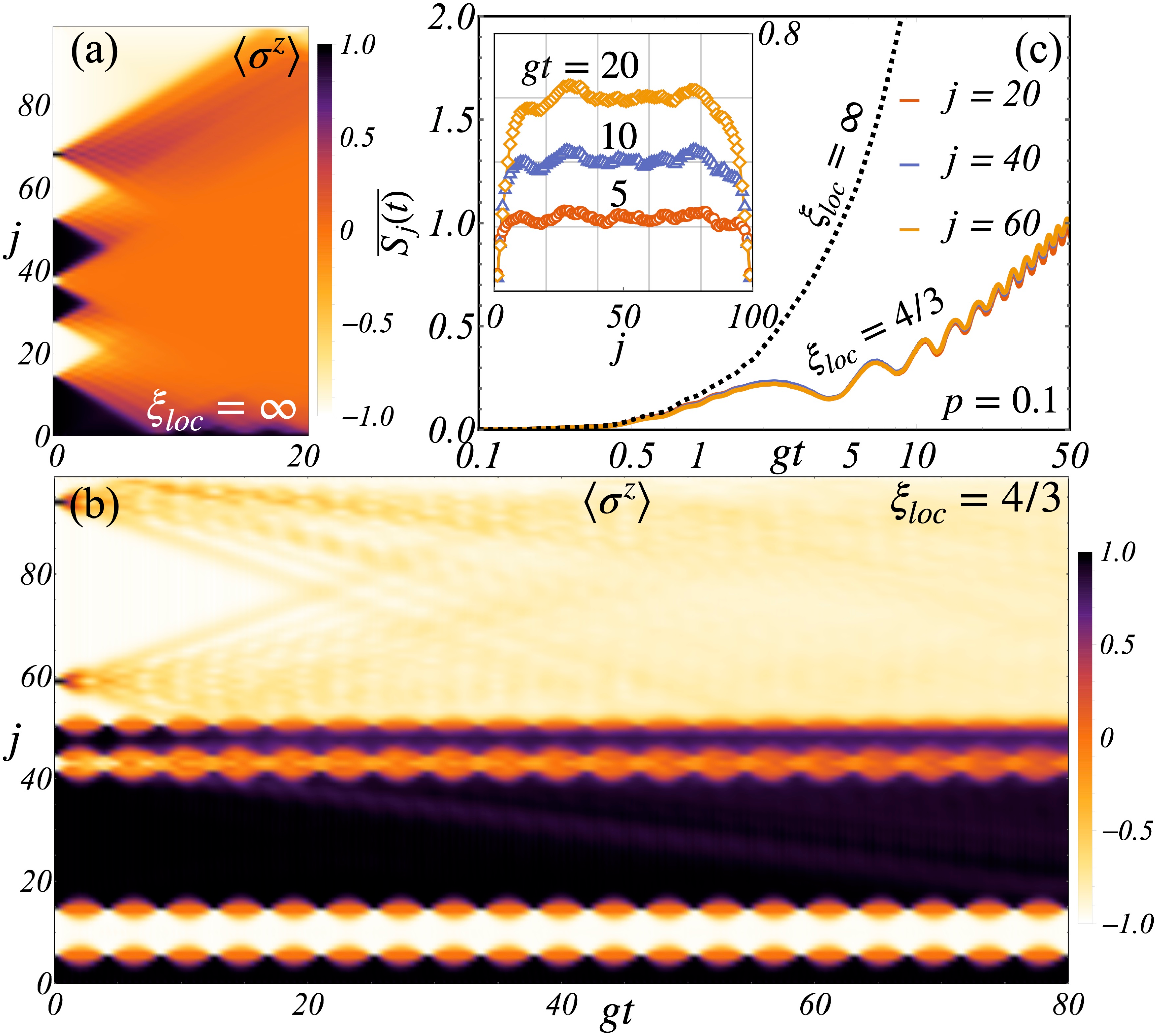}
\caption{%
\textbf{Effects of confinement}: nonequilibrium evolution of the magnetization profile [panels (a) and (b)] and of entanglement (c) in a quantum Ising chain. 
$L=100$ spins are initialized in a random product state 
with a density $p=0.1$ of longitudinal domain-walls. {The quantum evolution is simulated via the time-evolving block-decimation algorithm on matrix-product states with maximum bond dimension D=300  \cite{VidalTEBD}.} 
The dynamics are generated by $H$ in Eq.~\eqref{eq:Ising} with $J=5g$, and
(a) $h=0$: in the absence of confinement, domain-walls freely propagate, 
smoothening out all spatial inhomogeneities;
(b) $h=0.75g$: 
while confined bound states of closeby domain-walls diffuse (upper half of the plot), 
isolated domain-walls are 
Stark-localized by linear confining potentials, and perform coherent Bloch oscillations of spatial amplitude $\xi_{\text{loc}}=g/h$ (lower half of the plot). 
Panel
(c): dynamics of the von Neumann entanglement entropy $S_j(t)$ for different position $j$ of the bipartition cut, averaged over $500$ initial states. 
$S_j(t)$ grows linearly in the deconfined limit (a), $\xi_{\text{loc}}=\infty$ 
and logarithmically in the presence of confinement (b), $\xi_{\text{loc}}=4/3$, 
as also emphasized by the  inset.
These qualitative features are unaltered upon
 varying the localization length $\xi_{\text{loc}}$ while keeping $p \lesssim 1/(2\xi_{\text{loc}})$ and  $J \gg \abs{g},\abs{h}$.  
}
\label{fig_1}
\end{figure}
\noindent 
Remarkably, these two phenomenona stabilize nonthermal behavior and low entanglement for extremely long times in a thermodynamically relevant portion of the many-body Hilbert space, as illustrated in Fig. \ref{fig_1}.

\emph{Confinement and gauge invariance in one dimension.}—
The  occurrence of the phenomena mentioned above relies solely on the presence of confinement, and hence they emerge in both lattice gauge theories (LGTs) and {statistical-physics models} such as quantum spin chains. An exact correspondence between the two can be formulated in one spatial dimension, via the introduction/elimination of ancillary degrees of freedom together with local dynamical constraints. 
This leads to a unified framework  
for this broad class of systems \cite{McCoyIsingLGT,ShankarConfinement}. 
In this article, for concreteness, we focus on the paradigmatic quantum Ising chain. To illustrate the general equivalence above, we show how  this model can be exactly mapped onto a LGT with local $U(1)$ symmetry.
In the Supplemental Material \cite{SM}, 
we  consider other models with confined excitations, including the 
lattice Schwinger model of quantum electrodynamics \cite{SchwingerModel,Wilson74,KogutSusskindFormulation,KogutReviewLGT,Banks1976} and the antiferromagnetic XXZ spin chain in a staggered 
field, which describes certain anisotropic magnetic insulators 
\cite{RutkevichConfinementXXZ,WangSpinonConfinement,BeraEsslerSpinonConfinement,YangYang,wang2018experimental}.

\begin{figure}[t]
\includegraphics[width=0.48\textwidth]{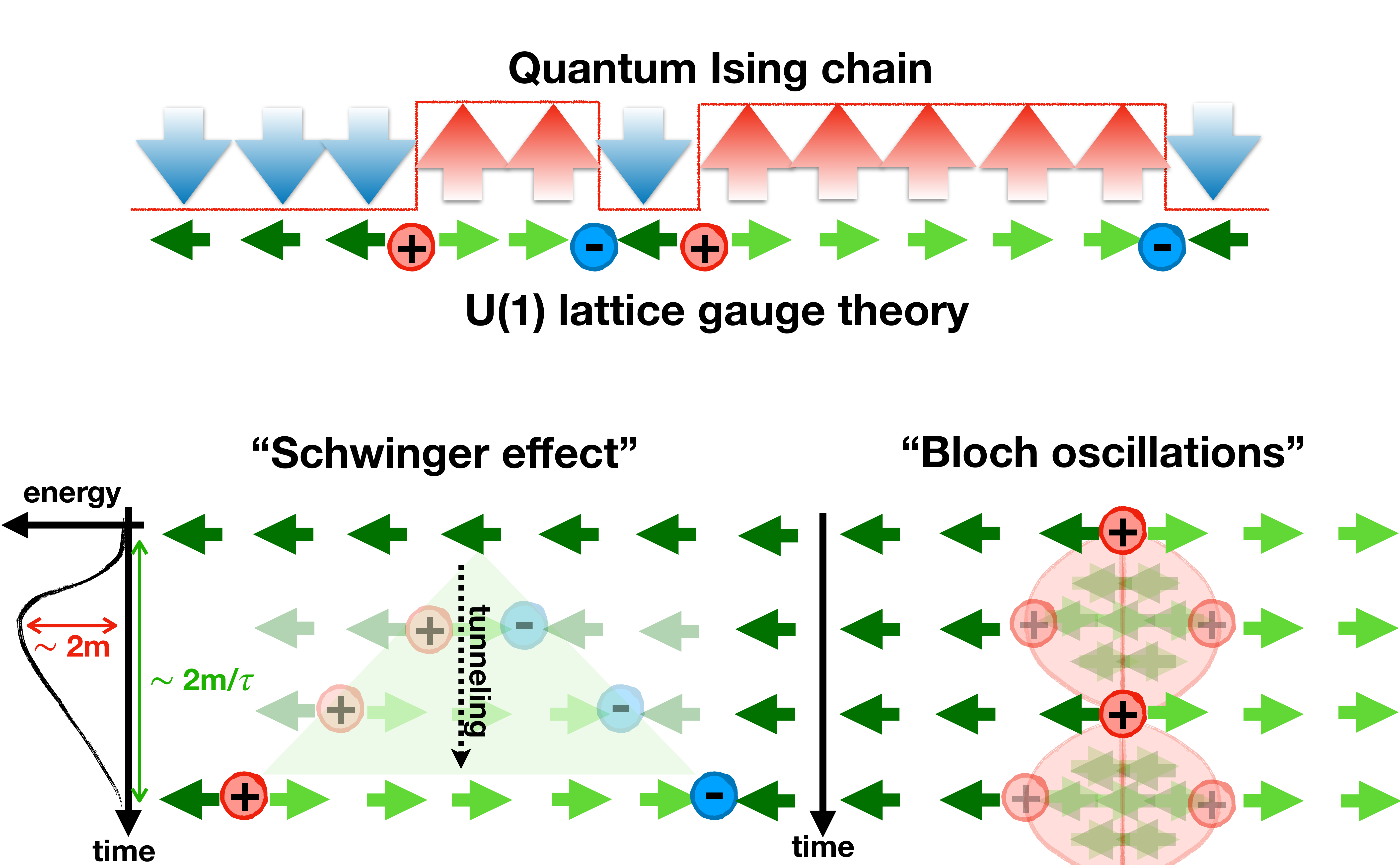}
\caption{\textbf{Mapping between a quantum spin chain and a LGT:}
Cartoon of
the mapping of the quantum Ising chain in Eq. \eqref{eq:Ising} onto the $1+1$-dimensional $U(1)$ lattice gauge theory in Eq. \eqref{HU1} (top), and of the two key mechanisms which render the resulting dynamics slow: 
suppression of false vacuum decay for weak coupling ("Schwinger effect", bottom left), and
Stark-localization of particles in a linear potential ("Bloch oscillations", bottom right). 
}
\label{fig_2}
\end{figure}

The quantum Ising chain is defined by the Hamiltonian
\beq
H= - J \sum_{j=1}^{L-1} \sigma^z_j \sigma^z_{j+1} - h \sum_{j=1}^L \sigma^z_j - g \sum_{j=1}^L \sigma^x_j ,
\label{eq:Ising}
\eeq
where $\sigma^{x,y,z}_j$, are Pauli matrices acting on site $j$.
The correspondence is based on the interpretation of the spin polarization operator $s^z_j\equiv \sigma^z_j/2$  
as a local "electric flux".  
{Fictitious fermionic matter degrees of freedom are introduced on the sites of the dual chain (i.e., on the bonds of the original chain 
\cite{KogutReviewLGT,susskind1977lattice}\footnote{Note that our convenient choice is opposite to Wilson prescription for lattice regularization of gauge theories, whereby matter and gauge degrees of freedom are placed on lattice sites and bonds respectively. In one-dimensional chains, though, sites and bonds are actually interchangeable. }%
): they represent "positrons" and "electrons".
Thus, crucially, one enforces local dynamical constraints that associate a kink (antikink) in the spin configuration with the presence of a positron (electron) on the corresponding bond, as described in Fig. \ref{fig_2}, top panel. 
These constraints are interpreted as implementing a discrete Gauss law and result from the $U(1)$ gauge-invariance of matter-field interactions.} 
%
%

To make this construction explicit, we define two species of fermions, positively ($p$) and negatively ($e$) charged respectively, residing on the chain bonds (denoted as half-integer sites), with corresponding creation operators $(c^{p,e}_{j+1/2})^\dagger$ 
and occupation numbers $n_{j+1/2}^{p,e} = (c_{j+1/2}^{p,e})^\dagger c_{j+1/2}^{p,e}$. 
We introduce 
a spin-$1/2$ $U(1)$-quantum link model \cite{Chandrasekharan1997,QLink1},
\beq
\label{HU1}
H_{U(1)} = H_{\rm{m}} + H_{\rm{g}} + H_{\rm{int}} \, ,
\eeq
with
\begin{align}
H_{\rm{m}} = m \sum_j
& 
(n_{j+1/2}^p + n_{j+1/2}^e) + U \sum_j n_{j+1/2}^p n_{j+1/2}^e \, ,
\nonumber \\
H_{\rm{g}} = \frac{\tau}{2}  \sum_j 
&
\sigma^z_j \, ,  
\nonumber   \\
H_{\rm{int}} = \nonumber
 w \sum_j & 
\Big\{ \big[
(c^p_{j-1/2})^\dagger +c^e_{j-1/2}\big] \sigma^+_j  \big[c^p_{j+1/2}+(c^e_{j+1/2})^\dagger\big] \\ &+ \text{h.c.}  \nonumber
   \Big\} \,   , 
\end{align}
where $\sigma^{\pm}_j = (\sigma^x_j \pm i \sigma^y_j)/2$ act as $U(1)$ parallel transporters \cite{KogutReviewLGT}. 
$H_{\rm{m}}$ encodes the fermion mass and onsite Hubbard-like interaction, and
$H_{\rm{g}}$ can be interpreted as the energy shift caused by a background field (or topological $\theta$-angle \cite{Coleman1976239,SuraceRydberg}).
In $H_{\rm{int}}$, the various terms describe hopping and pair creation/annihilation of fermions. 
The $U(1)$ gauge invariance of these interactions is expressed by the local symmetries 
$
[H,G_{j+1/2}] =0
$
with 
$
G_{j+1/2} =
\sigma^z_{j+1}/2 - \sigma^z_{j}/2
- (n_{j+1/2}^p - n_{j+1/2}^e )
$.
{
Accordingly, the complete Hilbert space decomposes into dynamically disconnected subspaces, labelled by the set of eigenvalues $\{q_{j+1/2}=0,\pm 1,\pm 2\}$ of $\{G_{j+1/2}\}$, interpreted as static background charges.
Here we focus on the neutral gauge sector, i.e., on the space of the states $\ket{\psi}$ for which the Gauss law $G_{j+1/2} \ket{\psi}\equiv 0$ is satisfied at all sites $j$, i.e., $q_{j+1/2}\equiv0$ \cite{KogutReviewLGT}.}
This law asserts that the variation 
of the gauge field strength $\sigma^z/2$ upon crossing a bond $(j,j+1)$ equals the {dynamical} charge $Q_{j+1/2} = n^p_{j+1/2} - n^e_{j+1/2}$ located on it.

In the presence of a strong Hubbard repulsion $U\to\infty$, each "classical configuration"
of the gauge field (eigenstate of all $\sigma^z_j$ operators)  fully determines a unique configuration of the matter particles via the Gauss law. 
This allows one to eliminate the redundant matter degrees of freedom \cite{SuraceRydberg,ZoharCiracMatterIntegration} and write the model in terms of a \emph{locally} self-interacting gauge field \cite{Coleman1976239}.\footnote{We observe that the converse elimination of the gauge field is also possible in one dimension, but the resulting model of interacting matter particles involves long-range Coulomb forces.}
In this case, all matrix elements of the Hamiltonian \eqref{HU1} between two classical gauge-field configurations coincide with the corresponding matrix elements of the quantum Ising chain \review{[Eq. \eqref{eq:Ising}]} in the $\sigma^z$-basis, upon identifying $m=2J$, $\tau=-2 h$, $w= - g$, 
 and up to an overall energy shift (see the Supplemental Material \cite{SM} for details).
Within this LGT picture, the longitudinal field $h$ in the quantum Ising chain plays the role of the electrostatic string tension $\tau$, leading to particle confinement.
In passing, we mention that the Ising chain can also be mapped onto a $\mathbb{Z}_2$-LGT, see Ref. \cite{SM}.

{Below we will analyze in detail the quantum Ising chain, but the conclusions apply to the general class of quantum chains with confinement of excitations \cite{SM}.}

\emph{Exponential suppression of pair creation.}— 
When a particle and an antiparticle in the vacuum are adiabatically separated at a distance $d$, the energy $E(d)\sim \tau d$ associated with the gauge-field string linking them grows proportionally to $d$ and eventually it overcomes the threshold $ E_{\text{min}}\sim 2m$ for the creation of a 
new pair. 
We argue that the 
\emph{dynamical} breaking 
of strings after a quench of the interactions takes anomalously long times for large values of the mass.
The mechanism for this suppression
may be essentially understood 
as a tunneling process across a high energy barrier. 
In fact, the decay process which converts the large amount of potential energy stored in long gauge-field strings into the energy of additional particle-antiparticle pairs is energetically allowed and entropically favorable, because a string state is very atypical compared to many-particle states with the same total energy. 
Accordingly, thermalization requires string breaking.
However, due to the energy conservation, the created particle and antiparticle of a pair must be separated at such a distance $d$ that the energy $\tau d$ they subtract from the broken string portion equals their mass, i.e., $\tau d  \sim 2m$. If the string tension $\tau$ is small compared to  the particle mass $m$, local pair creation is not possible, and virtual particles have to tunnel across a distance $d \sim 2m/\tau \gg 1$ in order for the string to decay --- see the bottom left panel of Fig. \ref{fig_2} for an illustration (here the lattice spacing is the unit length). 
This occurs through increasingly high-order processes in the interactions, and hence the decay is extremely slow.

The above qualitative picture is made quantitative by constructing the effective Hamiltonian in perturbation theory in $1/m$. 
%
%
We formally split the Hamiltonian into the mass term $H_0$, possessing highly-degenerate blocks, and the rest $V$, which involves gauge field and interactions. 
$H_0$ defines sectors of the Hilbert space labelled by the number of particles and well-separated in energy.
$V$ may contain block-diagonal matrix elements $
H_1$, describing particle/antiparticle energy and motion, and block-off-diagonal ones $
R_1=V-H_1$, corresponding to particle-antiparticle pair creation or annihilation. 
The latter processes are eliminated through a unitary transformation $e^{S_1}$.
{For the quantum Ising chain, 
the resulting effective Hamiltonian is }
$
H_{\text{eff}}^{(1)}  =  
- J \sum_j  \sigma^z_j \sigma^z_{j+1}  
- h \sum_j  \sigma^z_j 
- g \sum_j  ( P^\uparrow_{j-1} \sigma^x_j P^\downarrow_{j+1} + P^\downarrow_{j-1} \sigma^x_j P^\uparrow_{j+1} )
$, where $P_j^{\uparrow}$ ($P_j^{\downarrow}$) projects onto the "up" ("down") state of the $j$-th spin along $z$.

This standard  
procedure \cite{FrohlichSchriefferWolff,LossSchriefferWolff,McDonaldSchiefferWolffHubbard} {(often termed Schrieffer-Wolff transformation)} can be carried out to any arbitrary order $n$ in perturbation theory:
{One introduces higher-order terms $S_2,S_3,\dots$ in the generator of 
the unitary transformation 
$e^{S_{\le n}}$, with ${S_{\le n}=-S_{\le n}^\dagger=S_1+S_2+\dots+S_n}$. These terms are determined order by order} in such a way that the transformed Hamiltonian commutes with $H_0$ up to the $n+1$-th power of the perturbation strength, i.e.,
$
H' = e^{S_{\le n}} H e^{-S_{\le n}} =  H_{\text{eff}}^{(n)} +  V_{> n}
$,
with ${H_{\text{eff}}^{(n)} \equiv H_0 + H_1 + \dots + H_n} $,
and $[H_j,H_0]=0$ {(see the Supplemental Material \cite{SM} for  details)}. The effective Hamiltonian $H_{\text{eff}}^{(n)}$ preserves the block-diagonal structure of $H_0$ and accounts for all transitions within each sector of $H_0$ occurring through up to $n$ 
intermediate transitions involving states in different blocks (\emph{virtual} particle pairs). 
%
The perturbative series generated by 
this transformation are generally divergent at finite energy density, pointing to an asymptotic hybridization of the various blocks and hence thermalization.  
However, by adapting the rigorous theory in Ref. \cite{AbaninRigorousPrethermalization} (see also the recent Ref. \cite{DeRoeck19_GeneralPrethermalization}), one finds that by truncating the series  at an ``optimal order'' $n^*$ that scales linearly with the particle mass $m$,
the rest $V_{>n^*}$ can be made \emph{exponentially} small in $m$. 
Consequently, the effect of the latter can be neglected for exponentially long times. 
Denoting $H_{\text{eff}}^{(n^*)} \equiv H_{\text{eff}}$ and $S_{\le n^*} \equiv S$,  
the nonequilibrium evolution of the system is accurately described by 
\beq
\Ket{\Psi(t)} \simeq e^{-S} \, e^{-itH_{\text{eff}}} \, e^{S} \Ket{\Psi(t=0)}.
\eeq
Within this transformed picture, 
the number of particles 
is exactly conserved by $H_{\text{eff}}$, and hence it is approximately conserved by $H$ in the original picture at least for exponentially long times.
{This 
implies the emergence of nonthermal behavior in highly-excited states, 
signalled by the time-evolution of local observables such as the particle mass density, the gauge field, and the energy density.
In fact, the analysis above shows that 
the bulk of a long gauge string is stable against pair creation, since the  "string-breaking" (or "vacuum-decay") time scale is exponentially long in $m$.
This bulk stability persists in the continuum limit 
\cite{RutkevichFalseVacuum}, and, within the mapping in Eq. \eqref{HU1}, it is reminiscent of the Schwinger effect in quantum electrodynamics \cite{SchwingerMechanism}, in that the decay rate $\Gamma(\mathcal{E})$ per unit volume of a false vacuum in the presence of a background electric field $\mathcal{E}$ into particle pairs, is exponentially small  in the ratio between the electron mass $m$ and the electrostatic energy $\abs{e\mathcal{E}} \times 1/m$ contained within a Compton length, i.e., 
$\Gamma(\mathcal{E}) \propto (e\mathcal{E})^2  \exp \big(  -\frac{\pi m^2}{\abs{ e \mathcal{E}}}  \big)$, where $e$ is the electron charge and $\hbar=c=1$ \cite{SchwingerMechanism,ItzyksonZuber}.

In the Supplemental Material \cite{SM}, we provide the details of the construction of $H_{\text{eff}}$ and discuss
the quantum Ising chain in Eq. \eqref{eq:Ising} and the lattice Schwinger model as specific cases.
In the former, for $J \gg \abs{g},\abs{h}$
the estimates adapted from Refs. \cite{AbaninRigorousPrethermalization,LinMotrunichExplicitQuasiconserved} lead to the quasiconservation of the spatial density of domain-walls  at times $t \ll T_{\text{sb}}$, where 
\beq 
T_{\text{sb}} \; \ge \; g^{-1} \exp \Big( \, \text{const} \times J/\sqrt{h^2+g^2} \, \Big) ,
\label{eq:Tsb}
\eeq
and the constant is independent of the parameters \cite{SM}.

\emph{Stark localization.}—
The nonequilibrium dynamics starting from a generic initial state may be expected to undergo \emph{prethermalization} to the Gibbs ensemble $ e^{-\beta H_{\text{eff}}}/Z$ defined by the effective (nonintegrable) Hamiltonian $H_{\text{eff}}$ discussed above, at the inverse temperature $\beta$ uniquely determined by the energy density of the initial state \cite{Langen2016,AbaninRigorousPrethermalization}.
Contrarily to this expectation, we demonstrate that the combination of confinement and lattice effects leads to a dramatic slowdown of prethermalization in a thermodynamically significant portion of the many-body Hilbert space.
{This phenomenon is due to the Stark localization of particles \cite{WannierStarkLocalization} in their mutual linear confining potential, which suppresses spatial propagation and energy transport for \emph{arbitrary} interaction strength.} 

We consider below many-particles states, with a diluteness parameter $p$, i.e., with an average separation of $1/p$ lattice sites between consecutive particles.
To disentangle the effect of having a finite particle mass --- leading to exponentially slow pair creation --- from the intrinsic slow dynamics of $H_{\text{eff}}$, 
{we analyze the nonequilibrium dynamics generated by the latter truncated at the lowest order.} 
The effective picture consists of a system of 
hopping hardcore particles in a constant electric field, subject to 
interactions. 
Higher-order terms in $H_{\text{eff}}$ do not alter the physics qualitatively, as they just renormalize the hopping amplitudes with small longer-range terms \cite{SM}. 

In the extremely dilute  limit  $p\ll 1$ the system consists of isolated particles moving in a linear potential, a so-called Wannier-Stark ladder. 
This problem can be solved exactly \cite{Fogedby1Kink}: eigenstates are product states of localized orbitals with equispaced energy levels $E_n\propto n$. 
For the quantum Ising chain in Eq.~\eqref{eq:Ising}, $E_n=2h n$ and the localized wavefunction centered around the site $n$ reads
$\Psi_j^{(n)} = \mathcal{J}_{n-j}(g/h) $,
where $\mathcal{J}_{\nu}$ is the  Bessel function of order $\nu$ \cite{SM}. The tails of this localized orbitals decay faster than exponentially for $|n-j|\gg g/h \equiv \xi_{\text{loc}}$. 
If the distance between consecutive particles is much larger than $\xi_{\text{loc}}$, transport and thermalization are suppressed, and particles perform coherent (Bloch) oscillations around their initial position, with spatial amplitude $\xi_{\text{loc}}$ and temporal period $\pi/h$ \cite{MazzaTransport}. 

However, delocalization gradually occurs as $\ell=1/p$ is made comparable with twice the localization length $2 \xi_{\text{loc}}$. To understand this phenomenon and the associated time scales, we consider an isolated string with a particle (kink) at site $n_1$ and an antiparticle (antikink) at site $n_2 > n_1$. 
%
In the center-of-mass frame described by the relative coordinate $n_- = n_2 - n_1>0$, {the problem reduces to a single-particle Wannier-Stark ladder with hopping $2 g \cos K$, where $K$ is the center-of-mass momentum, and subject to a hard wall at the origin, i.e., to the boundary condition $\psi_{n_-=0}\equiv 0$ \cite{SM,FogedbyTwoKinkSolution}. }
The solution consists of a discrete sequence of particle-antiparticle bound states ("mesons") labelled by $\ell=1,2,\dots$ with dispersion relations $E_\ell(K)$.
The wavefunctions $\Psi_{n_-}^{(\ell,K)} = \mathcal{J}_{\ell -n_-}(2 g \cos K /h)$ with $\ell \gg 2 \xi_{\text{loc}}$ are localized far away from the boundary $n_- =0 $: they are hardly affected by it, and hence their energy $E_\ell = 2 h \ell$ is independent of $K$. This implies that spatially extended bound states have asymptotically flat bands: the two particles perform uncorrelated Bloch oscillations at the edges of the string connecting them, while the quantum diffusion of their center of mass is suppressed. 
However, the presence of the boundary bends the dispersion relation $E_\ell(K)$ of bound states 
with an extension comparable to that of the Bloch oscillations. 
This leads to correlated ("rigid") motion of the string edges, and hence spatial delocalization and entanglement growth. 

{The 
correction $\delta E_\ell (K)$ to the energy level $E_\ell$ 
in the dilute regime $\ell \gtrsim 2 \xi_{\text{loc}}$
is found to be approximately
$-2g \cos K\, \mathcal J_\ell\left(2\xi_{\text{loc}}\cos K\right)\, \mathcal{J}_{\ell-1}\left(2\xi_{\text{loc}}\cos K\right)$  \cite{SM}.
The analysis of the resulting 
spreading velocities ${v^{\text{max}}_{\ell} = \Max_{K} \abs{\partial_K E_{\ell}(K)}}$ of bound states for varying quantum number $\ell$, 
leads to a sequence of delocalization time scales $T_{\text{dloc}}(\ell,\xi_{\text{loc}} )$
rapidly increasing as the ratio $\ell/\xi_{\text{loc}}$ increases;
 for large $\ell \gtrsim \xi_{\text{loc}}^2$, one has \cite{SM}
\beq 
T_{\text{dloc}}(\ell,\xi_{\text{loc}} ) \; \sim \;  
g^{-1} \,
(\ell!)^2  \ell^{-3/2} \; 
\xi_{\text{loc}}^{-2\ell+1}.
\label{eq:Tdel}
\eeq
As a result, the typical delocalization time scale is state-dependent  via the diluteness parameter $p$, unlike the string-breaking time scale $T_{\text{sb}}$ in Eq.~\eqref{eq:Tsb}.
We stress that the above equations are nonperturbative in $g/h=\xi_{\text{loc}}$ and hence valid for arbitrarily large values of this ratio.


\begin{figure}[t!]
\includegraphics[width=0.49\textwidth]{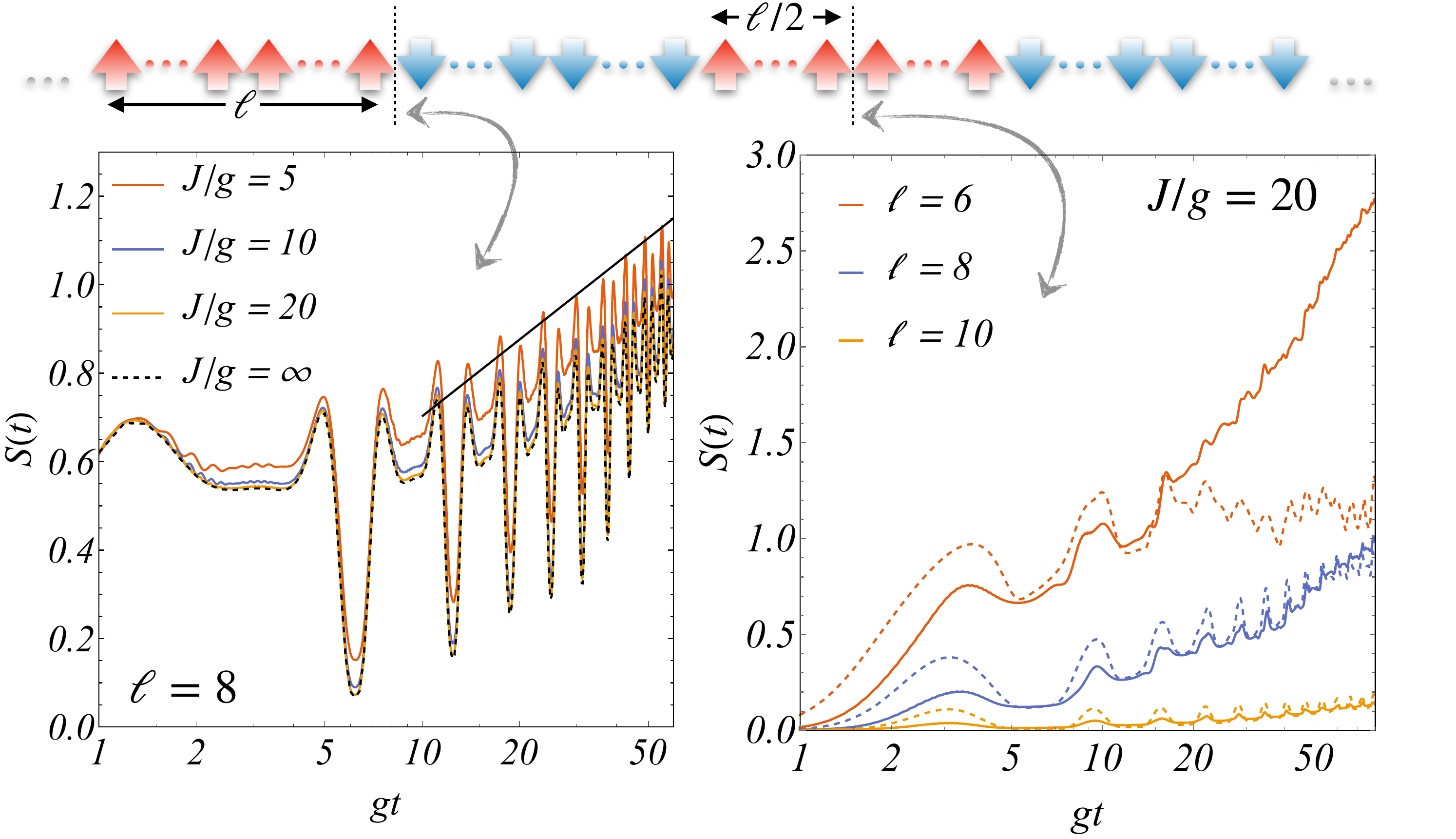}
\caption{\textbf{Signatures of slow dynamics:}
growth of the von Neumann entanglement entropy $S(t)$ 
{in the nonequilibrium dynamics of the quantum Ising chain in Eq. \eqref{eq:Ising}, numerically simulated via the TEBD algorithm},
starting from a state with equally spaced domain-walls at a distance $\ell$; the cartoon above the plots indicates the position of the bipartition cuts along the chain. 
Left: $S(t)$ exhibits pronounced coherent oscillations with frequency $ 2h$ superimposed to a slow growth (the straight line is a guide for the eye). Right: 
The growth of $S(t)$ slows down upon increasing the diluteness.
Dotted lines represent the growth of $S(t)$ in the evolution  of a single isolated string formed by the two domain-walls adjacent to the cut. The latter can be obtained analytically \cite{SM}, is upper-bounded by $ \log \ell + \text{const}$, and reaches its maximum around the time $T_{\text{dloc}}$, cf. Eq. \eqref{eq:Tdel}. 
Parameters: 
$\xi_{\text{loc}}=2$, $L=120$.
}
\label{fig_3}
\end{figure}

\emph{Slow entanglement growth.}— 
The  scenario outlined above sheds light on the effects of confinement on the nonequilibrium evolution of entanglement.
While the  entanglement entropy $S(t)$ is expected to increase linearly in time in generic quantum many-body systems which dynamically relax to equilibrium \cite{calabrese2004entanglement,CC_QuasiparticlePicture,FagottiCalabreseXY,
alba2018entanglement,kimhuse,nahum2017quantum,DeLucaChalker}, the quasilocalization discussed above is expected to cause a severe suppression of the growth of $S(t)$ despite the finite energy density, in analogy with disordered and glassy quantum systems \cite{HusePhenomenologyMBL,ProsenMBL,BardarsonPollmanMoore,Serbyn2013universal,Cirac:QuasiMBLWithoutDisorder,GarrahanMBLGlass,GarrahanQuasiMBLEast,AbaninProsen:SlowDynamics,SmithEE, SmithLong, brenes2018many}. 
This expectation is confirmed by numerical simulations using the time-evolving-block-decimation (TEBD) algorithm on matrix-product states \cite{VidalTEBD}, with maximum bond dimension $D=300$.
In particular, we initialize a quantum Ising chain of $L=100$ spins in nonentangled product states with a spatial density $p$ of domain-walls:
in Fig. \ref{fig_1} these states are drawn from a thermal ensemble $\rho_0 = e^{-\mu H_0}/Z$ of the "unperturbed" classical Ising chain with $p=[1-\tanh(\mu J)]/2$. 
(Similar dilute states with tuneable $p$ can be experimentally realized via the quantum Kibble-Zurek mechanism \cite{ZurekZoller,PolkovnikovNEQReview}).
The numerical results reported in Fig. \ref{fig_1} are compatible with a logarithmic growth of the bipartite entanglement entropy $S_j(t)$ superimposed to coherent oscillations of period $\pi/h$, ascribed to Bloch oscillations.
In Fig. \ref{fig_3}, instead, regularly arranged initial states are considered with equispaced domain-walls at a distance $\ell=1/p$ and $L=120$.
The fast convergence of $S(t)$ to that generated by the effective Hamiltonian $H_{\text{eff}}$ upon increasing $J$ (Fig. \ref{fig_3}, left panel) leads us to rule out the hypothesis that the slow vacuum decay is responsible for the entanglement growth.
Furthermore, the bottom right panel of Fig. \ref{fig_3} shows that the initial growth of $S(t)$ is captured by the delocalization of individual strings described in Eq. \eqref{eq:Tdel} above \cite{SM}; however, at longer times, many-particle effects lead to a slow unbounded growth.




\emph{Outlook.}— 
%
In the framework of localization phenomena in disorder-free quantum systems \cite{DeRoeckHuveneersScenario,DeRoeck1,Schiulaz1,Cirac:QuasiMBLWithoutDisorder,CarleoGlassy,BolsDeRoeckBoseHubbard,GarrahanMBLGlass,GarrahanQuasiMBLEast,AbaninProsen:SlowDynamics,QuasiMBLFrystration,RefaelStarkMBL,SchultzStarkMBL,GarrahanSlowDynamicsConstraints,PancottiQuantumEast}, this work establishes the role of confinement as a robust mechanism capable of dramatically slowing down the approach to equilibrium \cite{SmithDisorderFreeLocalization,SmithLong,BrenesConfinementLGT,
KCTC,MazzaTransport,SuraceRydberg,GorshkovConfinement,LeroseDWLR,Ercolessi2019, Tagliacozzo2019}. 
It is interesting to highlight the connection with the recently proposed "Stark many-body localization" 
\cite{RefaelStarkMBL,SchultzStarkMBL,KhemaniNandkishore_HigherD,Moudgalya_KrylovStark},
in that  the effective dynamics of the systems considered in the present work 
may be viewed 
as 
that of interacting particles in a constant field. 
Our preliminary numerical results suggest that rare high-density regions embedded in dilute systems do \emph{not} thermalize the rest of the system within the relevant time scales in this work;  
however, a complete analysis of this problem and of the various stages  of the dynamics \cite{Verdel19_ResonantSB}  calls for further investigations which we leave to future studies.

Our discussion applies to generic one-dimensional lattice models with confined excitations, including Abelian and non-Abelian LGTs \cite{Wilson74,PichlerMontangeroQLMTensorNetwork,kuhn2015}. 
The extension of our work to confining theories in higher dimensions stands as a challenging direction for future work, inasmuch as their real-time dynamics has hardly been explored in the framework of nonequilibrium statistical mechanics. \\



 \begin{acknowledgments}
 We thank F. Alet, R. Verdel, Y. Baum, J. Berges, W. De Roeck, E. Ercolessi, F. H. L. Essler, A. V. Gorshkov, M. Heyl, R. M. Konik, C. J. Lin, O. I. Motrunich, R. Nandkishore, E. van Nieuwenburg, S. Parameswaran, N. J. Robinson, A. Scardicchio, M. Schulz, J. Sisti, X. Turkeshi for interesting discussions and particularly D. A. Abanin, M. Dalmonte and M. M\"{u}ller for sharing with us their insights on the subject matter of this work. This work is partly supported by the ERC under grant number 758329 (AGEnTh).
 A. L., F. M. S. and P. P. M. thank the Galileo Galilei Institute for Theoretical Physics for the hospitality during the workshop \textit{"Breakdown Of Ergodicity In Isolated Quantum Systems: From Glassiness To Localization"} and the INFN for partial support during the completion of this work.
 \end{acknowledgments}

\bibliography{DWLR,EE,LGT,General}

\end{document}



\title{ Supplemental Material \\ {Quasilocalized} dynamics from confinement of quantum excitations \\  }

\author{Alessio Lerose}
\affiliation{SISSA -- International School for Advanced Studies, via Bonomea 265, 34136 Trieste, Italy.}
\affiliation{INFN, Sezione di Trieste, via Bonomea 265, 34136 Trieste, Italy.}
\author{Federica M.~Surace}
\affiliation{SISSA -- International School for Advanced Studies, via Bonomea 265, 34136 Trieste, Italy.}
\affiliation{ICTP -- International Center for Theoretical Physics, Strada Costiera 11, 34151 Trieste, Italy.}
\author{Paolo P. Mazza}
\affiliation{SISSA -- International School for Advanced Studies, via Bonomea 265, 34136 Trieste, Italy.}
\affiliation{INFN, Sezione di Trieste, via Bonomea 265, 34136 Trieste, Italy.}
\author{Gabriele Perfetto}
\affiliation{SISSA -- International School for Advanced Studies, via Bonomea 265, 34136 Trieste, Italy.}
\affiliation{INFN, Sezione di Trieste, via Bonomea 265, 34136 Trieste, Italy.}
\author{Mario Collura}
\affiliation{SISSA -- International School for Advanced Studies, via Bonomea 265, 34136 Trieste, Italy.}
\author{Andrea Gambassi}
\affiliation{SISSA -- International School for Advanced Studies, via Bonomea 265, 34136 Trieste, Italy.}
\affiliation{INFN, Sezione di Trieste, via Bonomea 265, 34136 Trieste, Italy.}
\maketitle

\setcounter{equation}{0}
\setcounter{figure}{0}
\setcounter{table}{0}
\setcounter{page}{1}
\makeatletter
\renewcommand{\theequation}{S\arabic{equation}}
\renewcommand{\thefigure}{S\arabic{figure}}
\renewcommand{\bibnumfmt}[1]{[S#1]}
\renewcommand{\citenumfont}[1]{S#1}

{Here} we derive in detail the results presented in the main text and we provide additional numerical evidence which demonstrates the occurrence of similar phenomena in other models with confined excitations. {The supplemental material is organized as follows:}  
in Section \ref{sec:mapping} we show how to map the quantum Ising chain onto a $U(1)$ or a $\mathbb{Z}_2$ lattice gauge theory (LGT);
in Section \ref{sec:SW_Ising} we discuss the general construction of the effective Hamiltonian and we report its analytic determination at the lowest order in perturbation theory in two cases: the quantum Ising chain and the lattice Schwinger model; 
in Section \ref{sec:twokinks} we study the effective model in the two-particle sector in order to estimate the delocalization time of an isolated string and rationalize the observed entanglement growth;
in Section \ref{sec_XXZ} we discuss the quasilocalized dynamics induced by confinement in the antiferromagnetic XXZ spin chain and in Section \ref{sec_LSM} in the lattice Schwinger model.

\section{Exact mapping between quantum spin chains and lattice gauge theories in one spatial dimension }
\label{sec:mapping}
In this Section, we provide the details of the mapping between quantum spin chains and one-dimensional LGTs.
The correspondence is based on the elimination of matter degrees of freedom. 
A related construction was recently proposed in Ref. \onlinecite{SuraceRydberg}, which allows one to interpret strongly interacting Rydberg atom arrays as the realization of a spin-$1/2$ $U(1)$
LGT with staggered fermionic matter.
For the sake of illustration, we focus here on the quantum Ising chain given in Eq.~(1) of the main text, but analogous mappings may be constructed for generic one-dimensional quantum lattice models by (i) introducing additional ``matter" degrees of freedom on the bonds, and (ii) defining gauge-invariant interactions in such a way that the Gauss law renders these newly introduced degrees of freedom actually redundant.

\subsection{The quantum Ising chain as a $U(1)$ LGT}

In the main text, we have argued that the quantum Ising chain can be mapped to a $U(1)$-LGT, with two species of fermions ("positrons" and "electrons"). In this Section, we detail the explicit mapping of the operators which allow one to transform the $U(1)$ lattice gauge theory in Eq.~(2) of the main text into the quantum Ising chain in Eq.~(1).

The first step consists in mapping the fermions $c^{p}_{j+1/2}$ and $c^{e}_{j+1/2}$ to hardcore bosons, by defining the Pauli spin-$1/2$ operators $\tau^\alpha_{j+1/2,\, p}$ and $\tau^\alpha_{j+1/2,\, e}$ as
\begin{align}
\label{eq_tauminusp}
&\tau_{j+1/2,\, p}^- = \prod_{k<j} \left[(-1)^{n_{k+1/2}^e}(-1)^{n_{k+1/2}^p}\sigma_k^z \sigma_{k+1}^z \right]c_{j+1/2}^p \, ,
\hspace{0.5cm}&
\tau_{j+1/2,\, p}^z= 2n_{j+1/2}^p-1
\, , \\
\label{eq_tauminuse}
&\tau_{j+1/2,\, e}^- = \prod_{k<j} \left[(-1)^{n_{k+1/2}^e}(-1)^{n_{k+1/2}^p}\sigma_k^z \sigma_{k+1}^z \right] (-1)^{n_{j+1/2}^e}(-1)^{n_{j+1/2}^p}\sigma_j^z \sigma_{j+1}^z c_{j+1/2}^e   \, ,
\hspace{0.5cm}&
\tau_{j+1/2,\, e}^z= 2n_{j+1/2}^e-1 \, ,
\end{align}
and $\tau_{j+1/2,\, p}^+=(\tau_{j+1/2,\, p}^-)^\dagger$, $\tau_{j+1/2,\, e}^+=(\tau_{j+1/2,\, e}^-)^\dagger$.
By exponentiating the Gauss law introduced in the main text, we find that in the gauge-invariant subspace one has
\beq (-1)^{G_j}= 
(-1)^{n_{j+1/2}^e}(-1)^{n_{j+1/2}^p}\sigma_j^z \sigma_{j+1}^z \equiv 1
\eeq
for every $j$. 
Plugging this relation into Eqs.~\eqref{eq_tauminusp} and \eqref{eq_tauminuse} we get that the equivalences $\tau_{j+1/2,\, p}^- = c_{j+1/2}^p$ and $\tau_{j+1/2,\, e}^- = c_{j+1/2}^e$ hold in this subspace. Note that in this step, we have been able to cancel the string coming from the Jordan-Wigner transformation by exploiting only the fact that $\mathbb{Z}_2$ is a normal subgroup of $U(1)$~\cite{ZoharRemovingFermions}: this method is quite general and can be applied in any number of spatial dimensions.
After this procedure, the Gauss law takes the form $G_j = (\sigma_{j+1}^z-\sigma_{j}^z-\tau_{j+1/2,\, p}^z+\tau_{j+1/2,\, e}^z)/2=0$. 
In addition, the constraint given by the infinite Hubbard interaction (Eq.~(2) of the main text) excludes the state $\ket{\uparrow \uparrow}_{j+1/2}$ (where the first and second spins refer  to the eigenvectors of $\tau_{j+1/2,\, p}^z$ and $\tau_{j+1/2,\, e}^z$, respectively). 

We will use both the Gauss law and the aforementioned constraint to perform the next step of the mapping, which is the elimination of the matter degrees of freedom (the general procedure can be found in Ref.~\onlinecite{ZoharRemovingFermions}).
It is useful to work in the basis of the eigenstates of $\sigma^z$ and $\tau^z$ operators, where both constraints are diagonal. The basic observation is that, for a given spin configuration of the gauge fields, the state of the matter particles is uniquely fixed by Gauss law, with the following rules:

\beq
\left\{
\begin{array}{ccc}
     \sigma_{j+1}^z=+1,\sigma_{j}^z = -1 & \rightarrow & \tau_{j+1/2,\, p}^z=+1, \tau_{j+1/2,\, e}^z=-1 \, ,\\
     \sigma_{j+1}^z=\sigma_{j}^z & \rightarrow & \tau_{j+1/2,\, p}^z=-1, \tau_{j+1/2,\, e}^z=-1  \, , \\
     \sigma_{j+1}^z=-1,\sigma_{j}^z = +1 & \rightarrow & \tau_{j+1/2,\, p}^z=-1, \tau_{j+1/2,\, e}^z=+1 \, .
     \label{eq:configurations}
\end{array}
\right.
\eeq
In the case $\sigma_{j+1}^z=\sigma_{j}^z$, the option $\tau_{j+1/2,\, p}^z=+1, \tau_{j+1/2,\, e}^z=+1$, allowed by the Gauss law, is excluded by the constraint given by the Hubbard interaction, thus ensuring that the mapping is one-to-one.
The fact that the configuration of matter particles is completely determined by that of the gauge field in the gauge-invariant sector can be reformulated as follows: We can find a unitary transformation $U=e^{iA}$ which maps each gauge-invariant state to a product state of a gauge field state and a single reference state of the matter field (e.g., the matter vacuum state $\ket{0}_{\text{m}}$, with $\tau^z_{j+1/2,\, p,e}\ket{0}_{\text{m}}=-\ket{0}_{\text{m}}$ for every $j$)\footnote{We note that the transformed states do {\it not} belong to the gauge invariant subspace, and that $U$ is not gauge invariant.}. This can be done for example via the Hermitian operator
\beq
A=\frac{\pi}{2}\sum_j \big( P_{j}^{\uparrow}\tau_{j+1/2,\, e}^x P_{j+1}^{\downarrow}+P_{j}^{\downarrow}\tau_{j+1/2,\, p}^x P_{j+1}^{\uparrow}  \big),
\eeq
where $P^\uparrow_j$ and $P^\downarrow_j$ are the projectors on the $\ket{\uparrow}_j$ and $\ket{\downarrow}_j$ states respectively. By using Eq.~\eqref{eq:configurations}, one can see that, on gauge invariant states, the action of $U$ consists in flipping all and only the $\tau^z$-spins 
in the state $\ket{\uparrow}$. Moreover, for each state of our basis, the gauge field part is left invariant by $U$.

The unitary transformation $U$ effectively eliminates the redundant matter degrees of freedom. In fact,
we can now define the transformed Hamiltonian $H'={}_{\text{m}}\braket{0|U H U^\dagger|0}_{\text{m}}$ which acts on the non-trivial part (the gauge-field configurations) of the transformed states. 
We apply the transformation to each term of Eq.~(2) of the main text.
The mass term $H_{\text{m}}$ can be transformed by noting that, on gauge-invariant states, $(n_{j+1/2}^p + n_{j+1/2}^e)=(1-\sigma^z_j\sigma^z_{j+1})/2$. Then, by using the fact that $U$ acts as the identity on the gauge field part for each state of our basis, we find that
\beq
H'_{\text{m}}=\frac{m}{2} \sum_j (1-\sigma^z_j\sigma^z_{j+1}),
\hspace{2cm}
H'_{\text{g}} = \frac{\tau}{2}  \sum_j \sigma^z_j,
\hspace{2cm}
H'_{\text{int}}=w \sum_j \sigma_j^x,
\eeq 
i.e., $H'$ is  a quantum Ising chain in a transverse and longitudinal field. In addition, this establishes the correspondence between the parameters of the LGT and those of the quantum Ising chain anticipated in the main text (see Eq.~(1)).

 {It is interesting to finally comment on the gauge-integrated version of the above lattice gauge theory, where the gauge field is eliminated by solving the Gauss law~\cite{Banks1976,SalaVariationalStudy}.
 In one spatial dimension, the result of this procedure is a model of charges interacting via long-range Coulomb potentials.
 In the specific case of the 
 $U(1)$ 
 LGT discussed above, the gauge-integrated model is equivalent to a model of charges subject to a constant electric field and to the constraint of sign alternation along the chain. 
 The latter makes the particles interacting, as made explicit by the strong  on-site Hubbard repulsion.
 The slow dynamics discussed in the main text can thus be connected with the recently proposed ``Stark many-body localization" of interacting charged particles in a strong field \cite{RefaelStarkMBL,SchultzStarkMBL}.
 }


\subsection{The quantum Ising chain as a $\mathbb{Z}_2$ LGT}

In the previous Section, we have shown that the one-dimensional quantum Ising model in a transverse and longitudinal field can be mapped to a $U(1)$ lattice gauge theory. This was done by identifying the local magnetization along $z$ with the electric field and the kinks/antikinks with the particles/antiparticles.

We now demonstrate that, in a similar spirit, the quantum Ising chain can also be mapped to a $\mathbb{Z}_2$ LGT. 
The main difference with respect to the $U(1)$ LGT is that we now use the same fermionic operators to designate both kinks and antikinks. We therefore introduce fermions on bonds, with creation operators $c_{j+1/2}^\dagger$ and occupancy number $n_{j+1/2}=c_{j+1/2}^\dagger c_{j+1/2}$, and define the following Hamiltonian
\beq
\label{HZ2}
H_{\mathbb{Z}_2} = H_{\text{m}} + H_{\text{g}} + H_{\text{int}}
\eeq
with
\beq
    H_{\text{m}}=m \sum_j n_{j+1/2},
    \hspace{1cm}
    H_{\text{g}} = \frac{\tau}{2}  \sum_j  \sigma^z_j,
    \hspace{1cm}
    H_{\text{int}} = w \sum_j 
(c_{j-1,j}^\dagger \sigma^x_j c_{j+1/2} + c_{j-1/2}^\dagger \sigma^x_j c_{j+1/2}^{\dagger} +\text{ h.c.}).
\eeq

The operators
\beq
G_j=\sigma_j^z \sigma_{j+1}^z(1-2n_{j+1/2})
\eeq
commute with $H_{\mathbb{Z}_2}$ and generate a local $\mathbb{Z}_2$ gauge symmetry. The Gauss law $G_j \equiv 1$ thus restricts the Hilbert space to the neutral gauge sector, meaning that, as anticipated, the allowed configurations are those in which fermions sit on all the bonds where a kink or an antikink is present in the gauge field configuration.
{Hence, an analogous procedure of matter-integration as that detailed above for the $U(1)$ LGT mapping, shows that $H_{\mathbb{Z}_2}$ in Eq. \eqref{HZ2} is equivalent to the quantum Ising chain.
It is interesting to observe that the $U(1)$ LGT mapping above straightforwardly reduces to the $\mathbb{Z}_2$ LGT one here upon identifying matter particles with positive and negative charge to a single fermionic species.}


\section{Construction of the effective Hamiltonian}

\label{sec:SW_Ising}

We outline here the general procedure for the construction of the effective Schrieffer-Wolff Hamiltonian, following closely Refs.~\onlinecite{McDonaldSchiefferWolffHubbard,AbaninRigorousPrethermalization,LinMotrunichExplicitQuasiconserved} to arbitrary order in perturbation theory. With the notation introduced in the main text, $H_0$ denotes the ``unperturbed" block-diagonal Hamiltonian
given by the mass term, and the remaining terms are collected in $V=H-H_0$.
As in the main text, we introduce the generators $S_1,\dots, S_n$ of the transformation which brings $H$ to the desired block-diagonal form up to the various orders in perturbation theory.

Explicitly, the terms of the effective Hamiltonian and the generator of the unitary transformation are defined order by order in perturbation theory via the following recursive algorithm. We define $V_1\equiv V$ and for $n\ge 2$,
\beq
V_n = \sum_{(k_1,\dots,k_p)\in [n]'}
\frac{1}{p!} [S_{k_1},[S_{k_2},\dots,[S_{k_p},H_0]...]] 
+\sum_{(k_1,\dots,k_p)\in [n-1]}
\frac{1}{p!} [S_{k_1},[S_{k_2},\dots,[S_{k_p},V]...]],
\eeq
where the summations run over the set $[m]$ of the ordered partitions $(k_1,\dots,k_p)$ of an integer $m$, i.e., $k_i\ge 1$ and $\sum_{i=1}^p k_i = m$, and the prime $[m]'$ excludes the trivial partition $(k_1=m)$ with $p=1$. 
The operator $V_n$ represents the effective perturbation at the $n$-th order, i.e., the term of order $n$ in the transformed Hamiltonian after eliminating all block-off-diagonal transitions up to the $n-1$-th order, i.e., $H' = H_0 + \dots+H_{n-1}+V_n+V_{>n}$. Just like in the first order, we split the perturbation into a block-diagonal and a block-off-diagonal term, $V_n \equiv H_n + R_n$. The former constitutes the $n$-th order correction to the effective Schrieffer-Wolff Hamiltonian $H_{\text{eff}}$, while the latter is eliminated by choosing $S_{n}$ in such a way that $[S_n,H_0]+R_n = 0$.

This construction is algorithmic and may be carried out directly in the thermodynamic limit, as it involves only the commutation of local operators. However, manual derivations are limited to the first few orders because the combinatorial complexity of the calculation increases rapidly with the perturbative order $n$.
The convergence properties of this kind of construction have been discussed  
in Ref.~\onlinecite{AbaninRigorousPrethermalization} in full generality, and later in Ref.~\onlinecite{LinMotrunichExplicitQuasiconserved} in a specific case. 
We observe that, differently from these works, the local density of the unperturbed Hamiltonian $H_0$ need not be a single-site operator for our purposes (for example, in the quantum Ising chain, $\sigma^z_j \sigma^z_{j+1}$ is not);  however, the formal construction in the mentioned works may be adapted to the present case. 
It is rigorously shown therein that the relative magnitude of the ``rest" $V_{>n}$ compared to $H_0$ has an upper bound proportional to $n!$ times the perturbation strength to the power $n$.
The perturbative series (presumably) diverges, pointing to an asymptotic mixing of the eigenstates among sectors and thermalization. However, truncation of the series to order $n$ leads to a bound for the size of the effective perturbation at the $n$-th step. The optimal order $n^*$ (the one which gives the tightest bound) scales as the inverse perturbation strength, which leads to an exponential bound. In the main text, we use this fact to prove that the effective Hamiltonian represents a good approximation for studying the dynamics up to times which become exponentially long upon increasing the inverse perturbation strength.

Below, we report and discuss the effective Hamiltonian for the quantum Ising chain and for the lattice Schwinger model calculated at the lowest orders in perturbation theory. 
In Sec. \ref{sec_XXZ}, we also discuss a prototypical condensed-matter model with confined excitations, namely the antiferromagnetic XXZ quantum spin chain in a staggered field.
We emphasize that its corresponding effective Hamiltonian  is analogous to those of the aforementioned models.

\subsection{Effective Hamiltonian of the quantum Ising chain}
\label{sec:Ising_effective_2}

\begin{figure}[h!]
    \centering
    \includegraphics[width=0.7\textwidth]{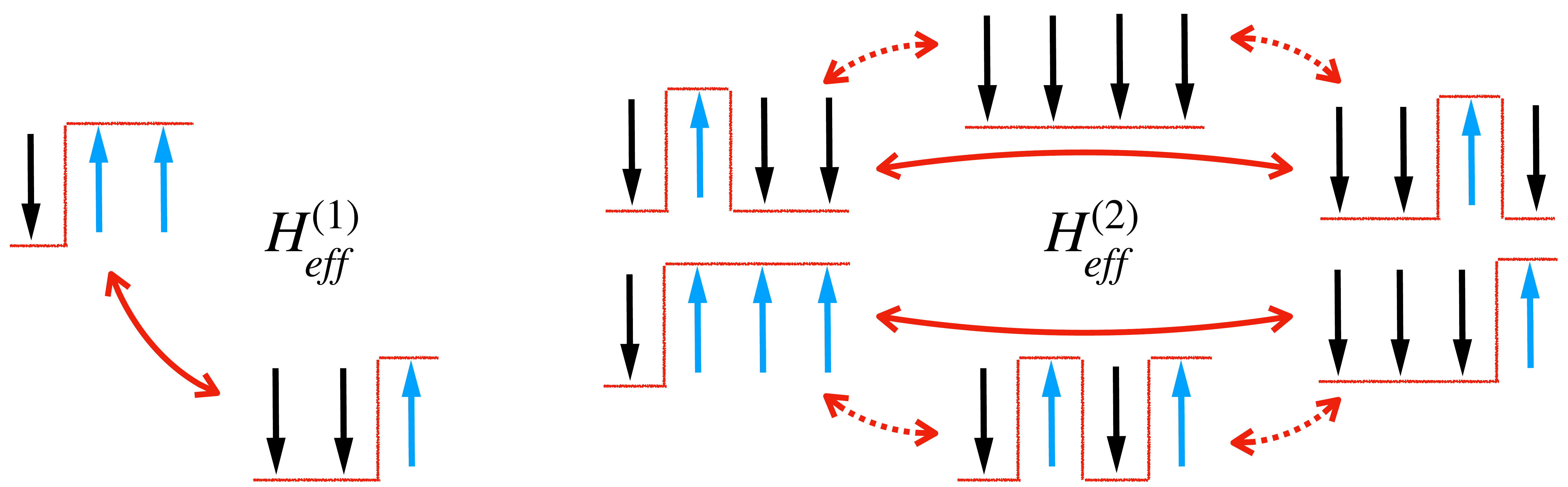}
    \caption{
    Cartoon of the perturbative transitions described by the effective Hamiltonian $H_{\text{eff}}^{(2)}$ of the quantum Ising chain up to the second order in $1/J$.
    %
At the first order, hopping of a kink/antikink by one lattice site is the only allowed transition. 
At the second order, one can either have hopping by one lattice site of a string/antistring of length one (top row) or hopping of two lattice sites of a kink/antikink (bottom row). 
Solid arrows show the block-diagonal transitions 
described by the effective Hamiltonian. 
The  intermediate states mediating the processes, indicated by dashed arrows,
involve ``virtual" states belonging to a different block.
The amplitudes of the second-order processes are proportional to $g^2/J$, see Eq. \eqref{eq:second_order_effective_Ising}.
    }
   \label{fig_QICtransitions}
\end{figure}

In order to show the structure of the effective Hamiltonian, we report its expression up to the third order for the quantum Ising chain in Eq.~(1) of the main text (using the notation introduced there):
\begin{align}
    H_0  = \;\;\;\;\; - J \sum_j & \; \sigma^z_j \sigma^z_{j+1}, \label{eq:zero_order_Ising} \\
    H_1  = \;\;\;\;\; - h \sum_j & \; \sigma^z_j - g \sum_j  \big( P^\uparrow_{j-1} \sigma^x_j P^\downarrow_{j+1} + P^\downarrow_{j-1} \sigma^x_j P^\uparrow_{j+1} \big) , \label{eq:first_order_Ising} \\
    H_2 = \;\; + \frac{g^2}{4J} \sum_j &
    \Big[ + P^\uparrow_{j-1} ( \sigma^-_j \sigma^+_{j+1}
    + \sigma^+_j \sigma^-_{j+1} )P^\uparrow_{j+2}   
    \label{eq:second_order_effective_Ising}
    + P^\downarrow_{j-1} ( \sigma^-_j \sigma^+_{j+1}
    + \sigma^+_j \sigma^-_{j+1} )P^\downarrow_{j+2}   \\
    & - P^\uparrow_{j-1} ( \sigma^+_j \sigma^+_{j+1}
    + \sigma^-_j \sigma^-_{j+1} )P^\downarrow_{j+2} 
    - P^\downarrow_{j-1} ( \sigma^+_j \sigma^+_{j+1}
    + \sigma^-_j \sigma^-_{j+1} )P^\uparrow_{j+2}  \nonumber \\
    & - \sigma^z_j \sigma^z_{j+1} \, \Big], \nonumber  \\
    H_3 =  +\frac{h g^2}{8J^2} \sum_j &
    \Big[
    -\sigma^z_{j} - \sigma^z_{j-1}\sigma^z_{j}\sigma^z_{j+1}
    \label{eq:third_order_Ising}
     \\ &
    + P^\uparrow_{j-1} ( \sigma^+_j \sigma^-_{j+1} + \sigma^-_j \sigma^+_{j+1} )  P^\uparrow_{j+2}
    - P^\downarrow_{j-1} ( \sigma^+_j \sigma^-_{j+1} + \sigma^-_j \sigma^+_{j+1} )  P^\downarrow_{j+2}
    \Big]
    \nonumber \\ 
     +\frac{ g^3}{8J^2} \sum_j &
    \Big[
    + P^\downarrow_{j-2} ( 
    \sigma^+_{j-1} \sigma^+_j \sigma^+_{j+1} +
    \sigma^-_{j-1} \sigma^-_j \sigma^-_{j+1} +
    \sigma^+_{j-1} \sigma^-_j \sigma^+_{j+1} +
    \sigma^-_{j-1} \sigma^+_j \sigma^-_{j+1} +
    )  P^\uparrow_{j+2}  \nonumber \\ &
    +  P^\uparrow_{j-2} ( 
    \sigma^+_{j-1} \sigma^+_j \sigma^+_{j+1} +
    \sigma^-_{j-1} \sigma^-_j \sigma^-_{j+1} +
    \sigma^+_{j-1} \sigma^-_j \sigma^+_{j+1} +
    \sigma^-_{j-1} \sigma^+_j \sigma^-_{j+1} +
    )  P^\downarrow_{j+2}  \nonumber \\ &
    -  P^\uparrow_{j-2} ( 
    \sigma^+_{j-1} \sigma^+_j \sigma^-_{j+1} +
    \sigma^-_{j-1} \sigma^+_j \sigma^+_{j+1} +
    \sigma^-_{j-1} \sigma^-_j \sigma^+_{j+1} +
    \sigma^+_{j-1} \sigma^-_j \sigma^-_{j+1} +
    )  P^\uparrow_{j+2}    \nonumber \\ &
     -  P^\downarrow_{j-2} ( 
    \sigma^+_{j-1} \sigma^+_j \sigma^-_{j+1} +
    \sigma^-_{j-1} \sigma^+_j \sigma^+_{j+1} +
    \sigma^-_{j-1} \sigma^-_j \sigma^+_{j+1} +
    \sigma^+_{j-1} \sigma^-_j \sigma^-_{j+1} +
    )  P^\downarrow_{j+2} \nonumber \\ &
    -  P^\uparrow_{j-1}  \sigma^x_j P^\downarrow_{j+1}
    -  P^\downarrow_{j-1}  \sigma^x_j P^\uparrow_{j+1}
    \Big] \, ,
    \nonumber
\end{align}
while the generators $S_1$ and $S_2$ of the unitary transformation up to the second order in $1/J$ are
\begin{align}
S_1 =  \frac{i g}{4 J}  \sum_{j} &
\big(
P^\uparrow_{j-1} \sigma^y_{j} P^\uparrow_{j+1} - P^\downarrow_{j-1} \sigma^y_{j}P^\downarrow_{j+1}
\big) \, ,  \\
S_2 = 
\frac{igh}{8 J^2}  
\sum_j & \big(
- 
P^\downarrow_{j-1} \sigma^y_{j} P^\downarrow_{j+1} -P^\uparrow_{j-1} \sigma^y_{j} P^\uparrow_{j+1} 
\big)
\\
     +\frac{g^2}{8 J^2}  \sum_j & \Big[ +P^\downarrow_{j-1} (\sigma^{-}_{j} \sigma^{-}_{j+1} - \sigma^{+}_{j} \sigma^{+}_{j+1}) P^\downarrow_{j+2} \nonumber \\ 
    & +P^\downarrow_{j-1}  (\sigma^{+}_{j} \sigma^{-}_{j+1}-\sigma^{-}_{j} \sigma^{+}_{j+1}) P^\uparrow_{j+2} \nonumber \\
    & +P^\uparrow_{j-1} (\sigma^{+}_{j} \sigma^{+}_{j+1}  -\sigma^{-}_{j} \sigma^{-}_{j+1}) P^\uparrow_{j+2} \nonumber  \\
    & +P^\uparrow_{j-1} (\sigma^{-}_{j} \sigma^{+}_{j+1} - \sigma^{+}_{j} \sigma^{-}_{j+1}) P^\downarrow_{j+2}  \Big] \, . \nonumber  \\
    \nonumber
\end{align}
One realizes that higher-order terms have a twofold effect:
they renormalize lower-order terms and introduce longer-range processes compatible with the conservation of $H_0$. Note that the maximal range of these processes at order $n$ is bounded by $n+2$, as can be proven by induction.
Transitions allowed up to the second order are sketched in Fig. \ref{fig_QICtransitions}.

For the quantum Ising chain, the estimates of Refs. \onlinecite{AbaninRigorousPrethermalization,LinMotrunichExplicitQuasiconserved} yield
\beq
\frac{\lvert \lvert V_{\ge n} \rvert\rvert}{\lvert \lvert H_0 \rvert\rvert} 
\; \le \; \bigg( \text{const } \times \;
n \, \frac{\sqrt{g^2+h^2}}{J}\bigg)^n ,
\eeq
where $\lvert \lvert \cdot \rvert \rvert$ indicates the operator norm of the local density of the argument.
Truncation of the series at the optimal order $n^* \propto J / \sqrt{g^2+h^2}$  leads to an exponential bound for the thermalization time, see Eq. (4) of the main text.

The construction presented here is similar to that 
of Ref.~\onlinecite{LinMotrunichOscillations} for the quantum Ising chain.
However, while that study is concerned with the homogeneous dynamics of elementary quasiparticle excitations above the ground state, we are here interested in the effective dynamics of dilute domain-walls, corresponding to high-energy states of the model.

\subsection{Effective Hamiltonian of the lattice Schwinger model}

In this Section we detail the construction of the effective Schrieffer-Wolff Hamiltonian for the lattice Schwinger model \cite{SchwingerModel,KogutSusskindFormulation} following the method of Sec.~\ref{sec:SW_Ising}. The latter represents the one-dimensional lattice discretization of quantum electrodynamics: matter particles are given by fermionic operators $\phi_j$ located at lattice site $j$ and interact via the $U(1)$ gauge electric field $E_{j,j+1}$ residing on the bonds of the lattice. The Hamiltonian reads
\begin{equation}
H = -w \sum_{j=1}^{L-1}(\phi^{\dagger}_j U_{j,j+1} \phi_{j+1} +\phi^{\dagger}_{j+1} U^{\dagger}_{j,j+1} \phi_{j}) + m \sum_{j=1}^{L}(-1)^j \phi^{\dagger}_{j} \phi_{j} + J \sum_{j=1}^{L} E_{j,j+1}^2     
\label{eq:Schwinger}
\end{equation}
where we adopted the Kogut-Susskind staggered fermion formulation \cite{KogutSusskindFormulation,susskind1977lattice}: particles occupying even sites represent quarks with positrons and empty odd sites represent electrons. The first term $w$ of the Hamiltonian gives the minimal coupling between gauge and matter degrees of freedom, the second is the particle/antiparticle mass $m$ and the last one is the electrostatic energy $J$. $U_{j,j+1}=\mbox{exp}(i \theta_{j,j+1})$ is the parallel transporter, with $\theta_{j,j+1}$ the vector potential, and it is related to the electric field $E_{j,j+1}$ by the commutation relations
\begin{equation}
[\theta_{j,j+1},E_{n,n+1}] = i \delta_{j,n}, \, \, \, \, [E_{j,j+1},U_{n,n+1}] = U_{j,j+1}\delta_{j,n}.   
\end{equation}
The infinitesimal generators of the local $U(1)$ symmetry are
\begin{equation}
G_j = E_{j,j+1}-E_{j-1,j} -\phi^{\dagger}_j \phi_j + \frac{1-(-1)^j}{2}, 
\end{equation}
therefore $[H,G_j]=0$ and we work in the neutral gauge sector where  no background charges $G_j |\psi \rangle = 0$ are present. The Gauss law is therefore satisfied at any lattice site and at any time during the dynamics.  

\begin{figure}[h!]
    \centering
    \includegraphics[width=0.7\textwidth]{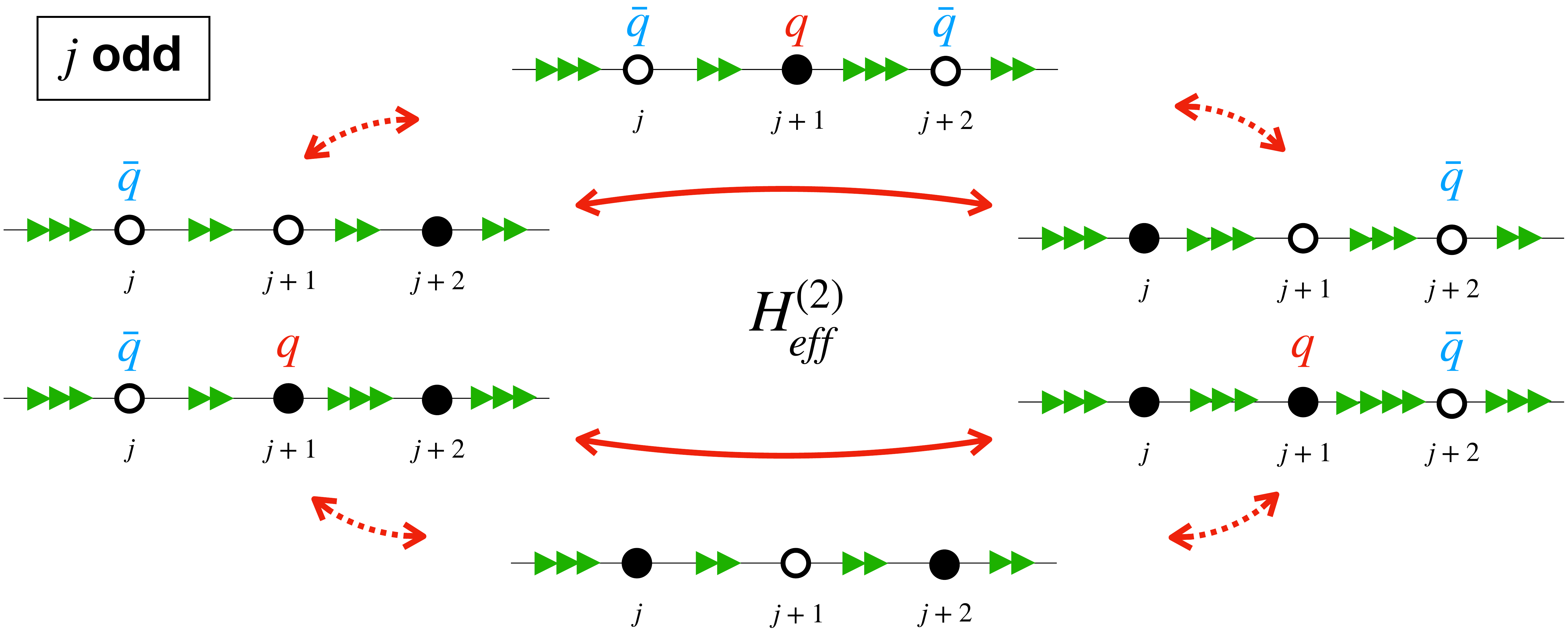}
    \caption{
    Sketch of the perturbative transitions described by the effective Hamiltonian $H_{\text{eff}}^{(2)}$ of the lattice Schwinger model in Eq.~\eqref{eq:effective_Hamiltonian_Schwinger} up to the second order in $1/m$.
    %
    With reference to the Hamiltonian $H$ in Eq. \eqref{eq:Schwinger}, black and white dots denote empty and occupied staggered-fermion sites, respectively; $q$  ($\bar{q}$) denote the presence of a  particle (antiparticle) at the corresponding site, with positive (negative) electric charge; the green arrows represent the value of the electric flux on the chain bonds.
    By virtue of the Gauss law, the electric flux jumps up (down) by one unit as a particle (antiparticle) is traversed from the left along the chain.
Considering for simplicity an odd lattice site $j$, one can either have hopping by two lattice sites of an anti-quark (top row) or hopping by two lattices of an anti-quark in presence of a quark occupying the site $j+1$ (bottom row). For even $j$ analogous processes take place, in which the role of the quark and the anti-quark are exchanged. 
Solid arrows show the block-diagonal transitions 
described by the effective Hamiltonian. The  intermediate states mediating the processes, indicated by dashed arrows, involve ``virtual" states belonging to a different block.
The amplitude of these transitions are proportional to $w^2/m$, see Eq. \eqref{eq:effective_Hamiltonian_Schwinger}.
    }
   \label{fig:Schwingertransitions}
\end{figure}

In order to derive the effective Hamiltonian governing the dynamics of the model in the limit of a large electron/positron mass $m$, we split the Hamiltonian of Eq.~\eqref{eq:Schwinger} as follows:
\begin{eqnarray}
H &=& H_0 + V, \\
H_0 &=& m \sum_{j=1}^{L}(-1)^j \phi^{\dagger}_{j} \phi_{j}, \label{eq:zero_order_mass_Schwinger} \\ 
V &=& H_1 + R_1 =  J \sum_{j=1}^{L} E_{j,j+1}^2 - w \sum_{j=1}^{L-1}(\phi^{\dagger}_j U_{j,j+1} \phi_{j+1} +\phi^{\dagger}_{j+1} U^{\dagger}_{j,j+1} \phi_{j}), \label{eq:perturbation_Schwinger}
\end{eqnarray}  
where the perturbation $V$ has been in turn decomposed into a diagonal part $V_{\text{diag}} \equiv H_1 $ given by the electrostatic term $J$ conserving the particle/antiparticle number (mass) and an off-diagonal one $V_{\text{offdiag}} \equiv R_1$, 
coupling sectors of the Hilbert space 
with different particle/anti-particle number (mass). 
By performing the unitary transformation as explained in Sec.~\ref{sec:SW_Ising}, the effective Hamiltonian $H_{\text{eff}}^{(2)}$,  which is 
block-diagonal up to second order in $1/m$, is found to be 
\begin{eqnarray}
H_{\text{eff}}^{(2)} &=& H_0 + H_1 + H_2, \label{eq:effective_Hamiltonian_Schwinger} \\ 
H_2 &=& \frac{w^2}{2m} \sum_{j=1}^{L}(-1)^j \phi^{\dagger}_j \phi_j + \frac{w^2}{2m} \sum_{j=1}^{L}(-1)^j (\phi^{\dagger}_j U_{j,j+1} U_{j+1,j+2} \phi_{j+2} + \phi^{\dagger}_{j+2} U^{\dagger}_{j+2,j+1} U^{\dagger}_{j+1,j} \phi_j), \label{eq:effective_H_2_Schwinger}
\end{eqnarray}
with $H_0$ given in Eq.~\eqref{eq:zero_order_mass_Schwinger}.
The lowest-order generators $S_1$ and $S_2$ of the unitary transformation $e^S$ bringing $H$ to the block-diagonal form $H_{\text{eff}}^{(2)}$ in Eq.~\eqref{eq:effective_H_2_Schwinger} may be written as
\begin{eqnarray}
S_1 &=& -\frac{w}{2m} \sum_{j=1}^{L} (-1)^j (\phi^{\dagger}_j U_{j,j+1}\phi_{j+1} -\phi^{\dagger}_{j+1} U^{\dagger}_{j,j+1} \phi_j),   \\
S_2 &=& \frac{w J}{4 m^2}  \sum_{j=1}^{L} \phi^{\dagger}_j(E_{j,j+1} U_{j,j+1}+U_{j,j+1}E_{j,j+1})\phi_{j+1} - \phi^{\dagger}_{j+1}(U^{\dagger}_{j,j+1}E_{j,j+1}+E_{j,j+1} U^{\dagger}_{j,j+1})\phi_j .    \\
\end{eqnarray}
The allowed processes at the second order in perturbation theory are described by $H_2$ and consist of a particle (anti-particle) hopping by two lattice sites mediated by a virtual state where a particle-antiparticle pair is either annihilated or created, as shown pictorially in Fig.~\ref{fig:Schwingertransitions}. 

We remark that in the lattice Schwinger model, the off-diagonal part $R_1$ of the perturbation in Eq.~\eqref{eq:perturbation_Schwinger} has no diagonal component and therefore the first non-trivial transitions  appear in the effective Hamiltonian at the second order. 
This differs from the case of the Ising chain in Sec.~\ref{sec:Ising_effective_2}, where the perturbation has a non-trivial diagonal component $H_1$ already at the first order, given in Eq.~\eqref{eq:first_order_Ising}.
This, in turn, implies that the dynamics in the lattice Schwinger model will be comparatively slower than that of the Ising chain.



\section{Two-particle problem and delocalization time}
\label{sec:twokinks}

In this Section we study the problem of a single string consisting of a particle at position $n_1$ and an antiparticle at position $n_2>n_1$. We consider the dynamics induced by $H_{\text{eff}}$, which conserves the number of particles and antiparticles. This two-body problem can be mapped to the problem of a single particle hopping on a two-dimensional lattice with coordinates $n_1$ and $n_2$ horizontally and vertically, confined to the half-plane $n_2>n_1$ by a hard-wall boundary condition along the diagonal and subject to a constant field orthogonal to the boundary.

We focus for simplicity on the quantum Ising chain, using the same notation as in Eq. (1) of the main text. Particles are here given by longitudinal domain-walls or kinks in the spin configuration, and
the lowest-order effective Hamiltonian in Eq. \eqref{eq:first_order_Ising} 
is the sum of $H_{\text{g}}$ and the projection of $H_{\text{int}}$; the mass term in Eq. \eqref{eq:zero_order_Ising} is constant within sectors and will be omitted. 
When restricted to the two-particle sector, 
 $H_{\text{g}}$ accounts for the linear confining potential between the two particles and $H_{\text{int}}$ for the particle hopping. 
Their respective matrix elements read
\beq
\braket{n_1, n_2|H_{\text{g}}|m_1, m_2}= 2h (n_2-n_1) \, \delta_{n_1, m_1}\delta_{n_2, m_2}
\eeq
and 
\beq
\braket{n_1, n_2|H_{\text{int}}|m_1, m_2}
=g (
\delta_{n_1+1, m_1} \delta_{n_2, m_2} +
\delta_{n_1-1, m_1} \delta_{n_2, m_2}
+\delta_{n_1, m_1} \delta_{n_2+1, m_2} +
\delta_{n_1, m_1} \delta_{n_2-1, m_2}
).
\eeq
We emphasize here that the analysis reported below is actually is \emph{nonperturbative} in the ratio $g/h$. 

Since the interaction part depends only on the positive distance $n_2-n_1$, it is convenient to consider the coordinates $n_\pm = n_2 \pm n_2$. 
Accordingly, by plugging the ansatz $\psi_{n_+,n_-}^{(K)}= e^{iKn_+} \psi_{n_-}^{(K)}$ into the Schr{\"{o}}dinger equation $H \psi = E \psi$, one realizes that
the plane wave $e^{iKn_+}$ factors out, and the problem reduces to the single-particle Wannier-Stark ladder 
\beq
2h n_- \psi_{n_-}^{(K)} + 2g \cos K \big[\psi_{n_- -1}^{(K)}+\psi_{n_- +1}^{(K)}\big] = E \psi_{n_-}^{(K)} \label{eq:single_string}
\eeq
subject to a hard wall at the origin $n_- = 0$, i.e., to the boundary condition $\psi_{n_-=0}^{(K)}\equiv 0$.

In the absence of the hard wall, the exact eigenfunctions are of the form $\Psi_{n_-}^{(j,K)}=\mathcal{J}_{j-n_-}(2\xi_{\text{loc}} \cos K)$ (where $\mathcal{J}_\nu$ is the Bessel function and $\xi_{\text{loc}}=g/h$), with energy $E_j=2hj$, independent of $K$. 
These wavefunctions decay faster than exponentially as the relative distance $n_-$ moves away from  $j>0$ by more than $2\xi_{\text{loc}}$ lattice spacings. 
Hence, for $j\gg 2\xi_{\text{loc}}$, the effect of the boundary condition is negiglible. 
The dispersion relations $E_j$ vs $K$ of bound states labelled by $j$ become completely flat in this limit: If the initial particles' wavefunction is concentrated on widely separated regions, their center of mass does not move and the two particles perform uncorrelated Bloch oscillations around their initial positions. The hard wall in $n_-=0$ is responsible for the failure of this occurrence, and its effect becomes manifest as $j$ approaches twice the extent $\xi_{\text{loc}}$ of the Bloch oscillations, such that the two particles' wavefunction tails overlap significantly.

\begin{figure}[t!]
\includegraphics[width=0.49\textwidth]{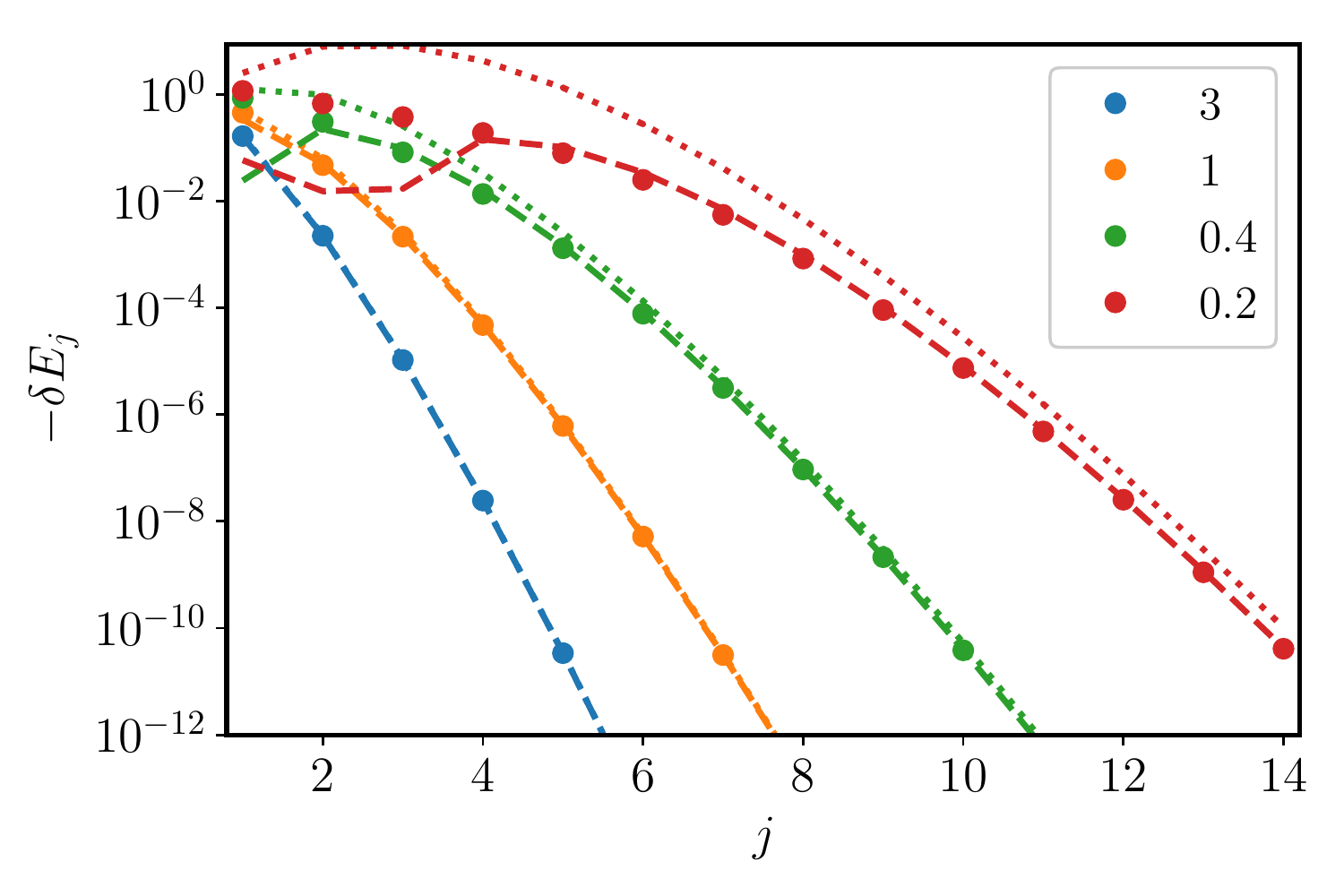}
\caption{Correction to the energy $E_j$ of the $j$-th eigenfunction, induced by the hard-wall potential. The dots correspond to the results of an exact diagonalization, the dashed line is the value estimated as $g\cos K \left[\left(\Psi_0^{(j,K)}\right)^*\Psi_1^{(j,K)}+\text{ h.c.}\right]$ and the dotted line is the one obtained from the last approximation in Eq.~\eqref{eq:correction}. The energies are in units of $g\cos K$ and the different colours refer to the different values of $h$ reported in the legend.
}
\label{fig:spectrum}
\end{figure}

We can quantify the resulting bending of the bands $E_j(K)$ by making an estimate based on the Hellmann-Feynman theorem. 
Consider the model without the hard wall, and let $2 \tilde g \cos K$ be the hopping amplitude between sites $n_-=0$ and $n_-=1$, and $2 g \cos K$ be the hopping amplitude on all the other links. Clearly, for $\tilde g=g$ the eigenfunctions and the eigenenergies are $\phi_{n_-}^{(j,K)}(\tilde g=g)=\Psi_{n_-}^{(j,K)}$, $E_j(K;\tilde g=g)=2hj$, respectively. 
On the other hand, if we adiabatically turn off $\tilde g$, we obtain the eigenfunctions  $\phi_{n_-}^{(j,K)}(\tilde g=0)$ and the eigenenergies $E_j(K;\tilde g=0)$ of the same model with the hard wall. From the Hellmann-Feynman theorem, we find that
\beq
\delta E_j(K) = E_j(K;\tilde g=0)-E_j(K;\tilde g=g)= 2\cos K \int_g^0  \left[\left(\phi_0^{(j,K)}(s)\right)^*\phi_1^{(j,K)}(s)+\text{ h.c.}\right] \mathrm{d}s.
\eeq
We can estimate this integral by replacing the integrand with the average between the values at the two extrema. Since $\phi_0^{(j,K)}(0)=0$, we find $\delta E_j \simeq - g\cos K \left[\left(\Psi_0^{(j,K)}\right)^*\Psi_1^{(j,K)}+\text{ h.c.}\right]$. 
We obtain the correction to the dispersion relation
\beq
\label{eq:correction}
\delta E_j(K) \simeq
-2g \cos K\, \mathcal J_j\left(2\xi_{\text{loc}}\cos K\right)\, \mathcal{J}_{j-1}\left(2\xi_{\text{loc}}\cos K\right)
\; \simeq \; -\frac{2\left(g\cos K\right)^{2j}}{j!(j-1)!\,h^{2j-1}} \, \, \, \,  , 
\eeq
where, in the last approximation, we used only the asymptotic behaviour of the eigenfunction for $j\gg  \xi_{\text{loc}}^2$ \cite{NIST:DLMF}.
In Fig.~\ref{fig:spectrum}, these approximations are compared with numerical diagonalization of the one-kink problem with the hard wall:
The first asymptotic estimate is found to be extremely accurate in the considered quasilocalized regime $j \gtrsim 2 \xi_{\text{loc}}$. 
In a semiclassical picture, asymptotically exact for $j\gg1$, $\xi_{\text{loc}}\gg1$ and fixed $j/\xi_{\text{loc}}$, this regime corresponds to having the two particles at the edges of the string performing \emph{non-overlapping} Bloch oscillations of amplitude $\xi_{\text{loc}}$ each.
We observe that the last asymptotic estimate in Eq. \eqref{eq:correction} agrees with the perturbation-theory argument for which nonvanishing corrections to the eigenenergy of the $j$-th bound state occur only at the $2j$-th order in $g/h$; we emphasize, however, that the equations above are valid for arbitrarily large localization lengths $\xi_{\text{loc}}=g/h$, provided the string length is even larger. 

The result in Eq.~\eqref{eq:correction} can be recovered from the exact single-string energy spectrum, expressed as a solution of the implicit equation
\begin{equation}
\mathcal{J}_{{-}E_j/2h}\left(2 \xi_{\text{loc}} \, \mbox{cos} K\right)=0,
\label{eq:exact_Fogedby_implicit}
\end{equation}
first derived in Ref.~\onlinecite{FogedbyTwoKinkSolution}. From the series expansion \cite{NIST:DLMF}
\begin{equation}
\mathcal{J}_{\nu}(z) = \left(\frac{z}{2}\right)^{\nu}   \sum_{p=0}^{\infty} \frac{(-1)^p}{\Gamma(p+1)\Gamma(p+1+\nu)}\left(\frac{z}{2} \right)^{2p} 
\end{equation}
one readily realizes that the leading correction $\delta E_j(K)$ in $z=2 \xi_{\text{loc}} \, \mbox{cos}K$ to the flat band energy level $E_j(K) = 2hj$ is given precisely by Eq.~\eqref{eq:correction}. 

From the above result, the maximal group velocity $v^{\text{max}}_{j} = \max_{K \in [0,{\pi})} \big|  \partial_K \, \delta E_j(K))\big|$ of the $j$-th bound state can be computed.
In particular, for $j\gg  \xi_{\text{loc}}^2$, one finds
\begin{equation}
v^{\text{max}}_{j}  \simeq \; h \frac{(2j)^{3/2}  }{(j!)^2} \left(\frac{g}{2h} \right)^{2j} \rm{e}^{-1/2} . 
\end{equation}
Note that these speeds dramatically drop to zero for $j \gg g/h$.
The delocalization time in Eq. (5) of the main text, is estimated by taking the inverse of $v^{\text{max}}_{\ell}$. 

We finally observe that from the exact eigenvalues and eigenfunctions of the two-kink problem, one can obtain the dynamics of the entanglement entropy $S(t)$ associated with a bipartition of the chain. 
The bottom right panel of Fig. 3 of the main text reports the growth of $S(t)$ for the evolution of isolated strings, compared with the corresponding growth for initial dilute states with multiple strings. 
For an initial condition given by kinks located at sites $i,j=n_{1,2}$ and bipartition cut at site $r$, the growth of $S(t)$  turns out to approximately consist of a discrete sequence of  "jumps", associated with the delocalization of the various components of the initial state on the eigenstates with quantum number $\ell=1,2,\dots$, their weight being maximal around $\ell \approx |n_2-n_1|$. 
Eventually, $S(t)$ converges to $\log 2$ as $t\to\infty$, since the diffusing string will asymptotically be either entirely on the left or entirely on the right of the cut, with equal amplitude.
Before this eventual saturation, $S(t)$ can attain values larger than $\log 2$, caused by transient correlations between the two particles located on opposite sides of the cut.
Using the fact that particles are confined,
it is straightforward to formulate an upper bound for $S(t)$.
In fact, for wavefunctions supported in the region $|i-j| \le d$, i.e., with the two particles separated by no more than $d$ lattice sites, the maximal von Neumann entanglement entropy is $\log (d+1)$.
For the considered initial condition, this bound holds with $d \approx |n_2-n_1|+2 \xi_{\text{loc}}$.

\section{Generality of the quasilocalized dynamics  }

In this Section, we provide numerical evidence of the fact that, as anticipated in the main text, most of the phenomena discussed here occurs generically in the presence of confinement. In particular, we consider here additional  one-dimensional models with confined excitations.

\subsection{The antiferromagnetic XXZ quantum spin chain in a staggered field}

\label{sec_XXZ}

\begin{figure}[h]
    \centering
    \includegraphics[scale=0.4]{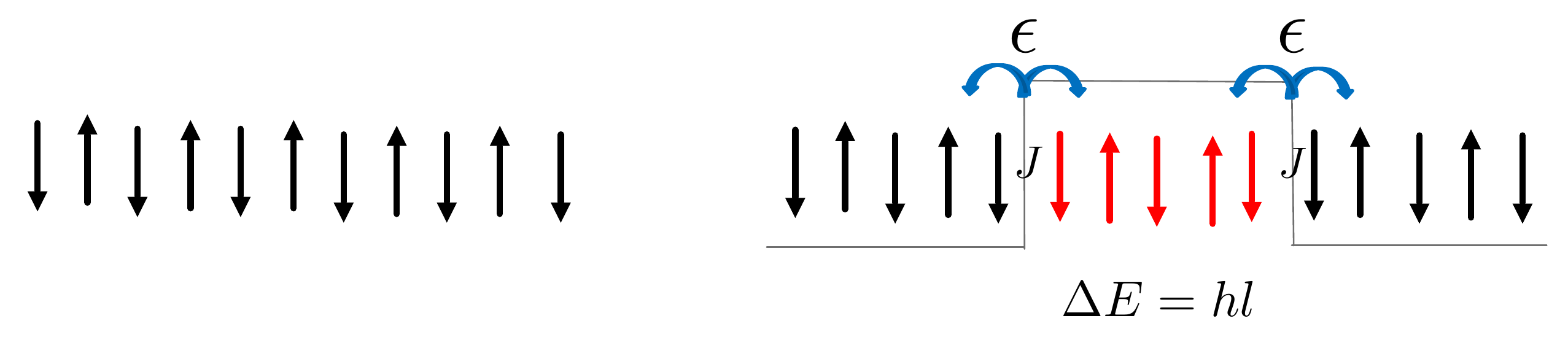}
    \caption{Sketch of the origin of spinon confinement in XXZ Hamiltonian in a staggered external magnetic field, described by Eq.~\eqref{Eq:XXZ_stag}.}
    \label{fig:Neel_sketch}
\end{figure}

As recalled in the introduction, confinement of excitations can be observed in a large class of one-dimensional statistical-physics models, and our findings apply to them.   
In this Section we corroborate this statement by studying the effects of confinement in the antiferromagnetic XXZ spin chains in a longitudinal staggered magnetic field, described by the Hamiltonian
\begin{equation}
    H_{XXZ}=\tilde{J}\sum_{i=1}^{L-1}(S_i^xS_{i+1}^x+S_i^yS_{i+1}^y+\Delta S_i^zS_{i+1}^z)-h\sum_{i=1}^L(-1)^iS^z_i,
    \label{Eq:XXZ_stag}
\end{equation}
where $S^\alpha_i=\frac{1}{2}\sigma^\alpha_i$. Let us note that in this case the magnetization is conserved: in fact, we have $$\Big[H_{XXZ}, {\sum_i} S^z_i\Big]=0.$$

The Hamiltonian \eqref{Eq:XXZ_stag} exhibits a quantum phase transition at $\Delta/\tilde{J} = 1$: For $\Delta/\tilde{J}>1$ the ground state has an antiferromagnetic axial order, whereas for $-1<\Delta/\tilde{J}<1$ it presents planar order. 
This model for $h=0$ was exactly solved by Yang and Yang using the Bethe ansatz~\cite{YangYang}. 
In the ordered phase 
the excitations can be described by kinks, as the Ising model in the ordered ferromagnetic phase ($h_z<J$). 
The ``Ising limit'' is, in fact, recovered when $|\Delta|\gg1$. 
The Hamiltonian~\eqref{Eq:XXZ_stag} describes with a good accuracy the intra-chain interactions in \ch{SrCo_2V_2O_8} crystals~\cite{Essler_conf,wang2018experimental}, in this case the external staggered magnetic field is given by the average effect of the inter-chains interactions. In Refs.~\onlinecite{RutkevichConfinementXXZ, Essler_conf}, the masses of the first bound states are computed, using semiclassical and perturbative approximations, and an almost perfect match is found  with the experimental data, obtained via neutron scattering. 
Here we are interested in the  dynamical properties of this model out of equilibrium. 
For this purpose we focus on the case $\Delta>1$, for which the ground state is N\'eel-ordered and gapped. 
In the limit $\Delta\gg1$ the twofold degenerate ground states are well approximated by the N\'eel ($|N\rangle$) and the anti-N\'eel ($|AN\rangle$) states
\begin{equation}
     |N\rangle=\bigotimes_{i=1}^{L/2}\upa_{2i} \dna_{{2i}+1}, \qquad |AN\rangle=\bigotimes_{i=1}^{L/2}\dna_{2i} \upa_{{2i}+1}.
\end{equation}
Kinks interpolate between these two vacua. These excitations are deconfined for $h=0$, i.e., they are not subject to long-range interactions
. 
If we switch on an external staggered magnetic field $h\neq 0$, the spectrum changes in a nonperturbative way. 
In particular, the two vacua are no longer degenerate: one is the true vacuum and the other the false vacuum lying in the middle of the spectrum. 
In this case the quasi-particle excitations feel a long-range confining potential, as sketched in Fig.~\ref{fig:Neel_sketch}.

\begin{figure}
    \centering
    \includegraphics[scale=0.4]{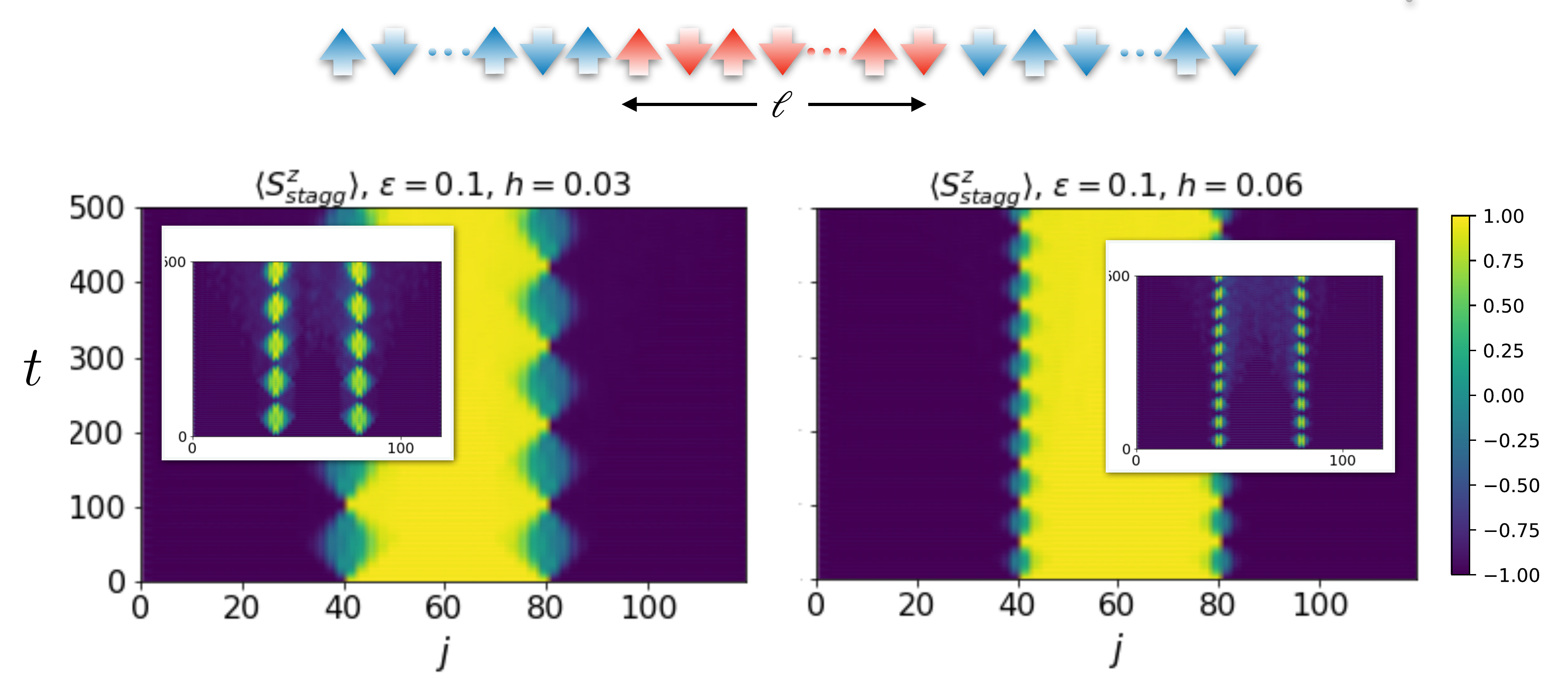}
    \caption{Evolution of the string state sketched in the cartoon below the plots, made up of two antiferromagnetic domain-walls, under the action of the Hamiltonian \eqref{Eq:XXZ_stag1}. 
    We can observe the stability of the string for different values of the external magnetic field. 
    The insets show the corresponding spatiotemporal variation of the bipartite entanglement entropy $S_{j}(t)$ betweens spins in regions $[1,j]$ and $[j+1,L]$. 
    In the bottom line we reported a sketch of the initial state.}
    \label{fig:XXZ_stag1}
\end{figure}
\begin{figure}
    \centering
    \includegraphics[scale=0.3]{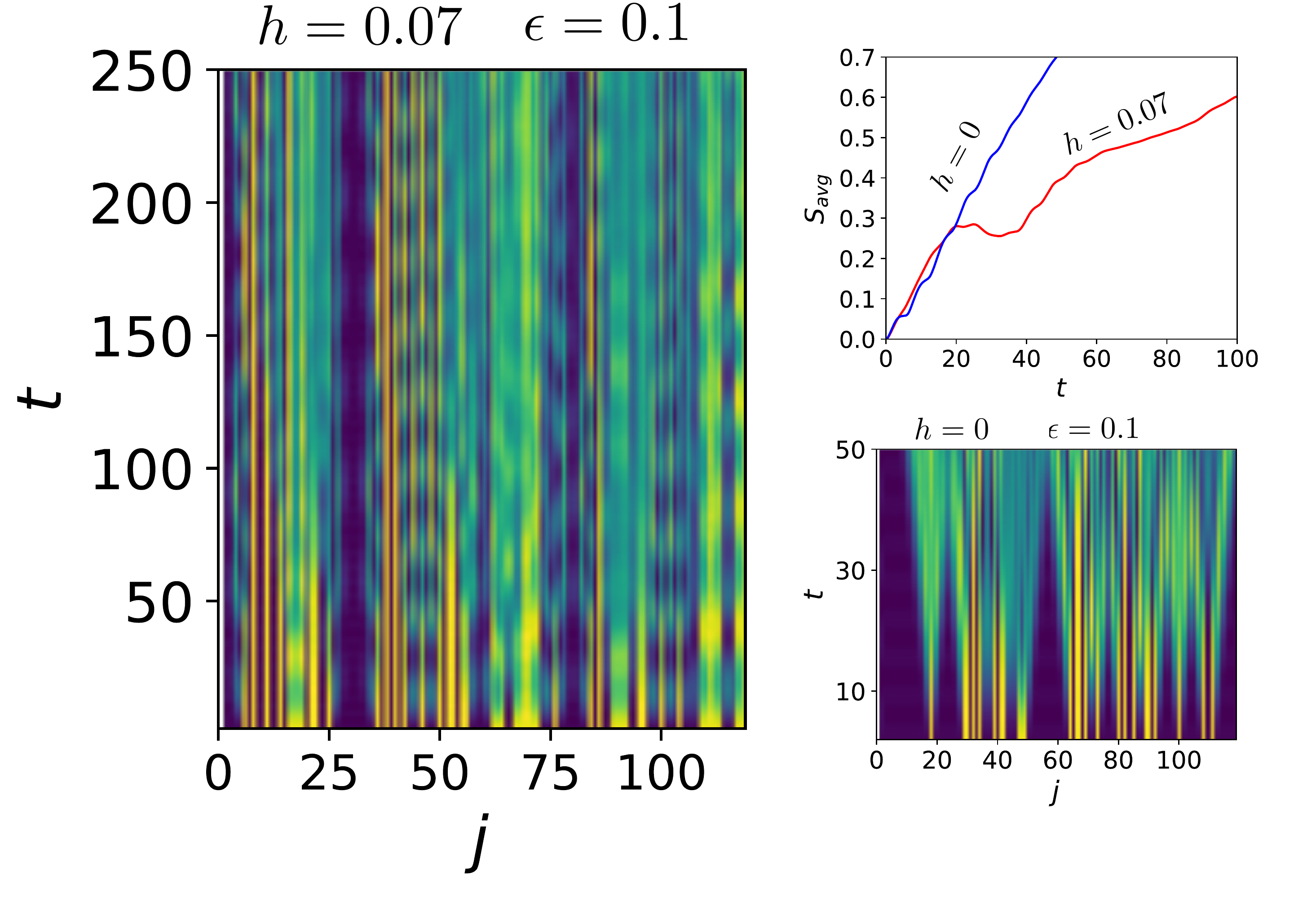}
    \caption{Density plot of the staggered magnetization for various values of the magnetic field and for a representative state with sizeable density $p=0.4$ of antiferromagnetic domain-walls. In particular, in the left panel $h=0.07$ and the initial inhomogeneity persists until long times. In the bottom right panel, instead, the same quantity is plotted for $h=0$, and one observes that the  initial inhomogeneities are smoothed out already at relatively short times. This qualitatively different behavior is signaled also by a different temporal growth of the  averaged entanglement entropy, reported in the top right panel.
    This figure should be compared with Fig. 1 of the main text.
    }
   \label{fig:XXZ_stag2}
\end{figure}

In order to study the dynamics, let us rewrite the Hamiltonian~\eqref{Eq:XXZ_stag}, defining $J=\tilde{J}/\Delta$ and $\epsilon=1/\Delta$, as
\begin{equation}
    H_{XXZ}=J\sum_{i=1}^{L-1}S_i^zS_{i+1}^z+\epsilon(S_i^xS_{i+1}^x+S_i^yS_{i+1}^y)-h\sum_{i=1}^L(-1)^iS^z_i.
    \label{Eq:XXZ_stag1}
\end{equation}
We fix $J=1$ and we consider $\epsilon<1$ in order to be in the anti-ferromagnetic phase. We consider both isolated strings and random initial states with multiple kinks, and we study the evolution of the staggered magnetization profile, i.e. $S^z_{j, stagg}=(-1)^jS_j^z$.
In Figs.~\ref{fig:XXZ_stag1} and  \ref{fig:XXZ_stag2} we report  the space-time density plot of the staggered magnetization starting from single string states and random states, respectively, with density $p=1/l$, where $l$ is the average separation between consecutive kinks along the chain. 
As we can observe, also in this case the dynamics exhibit signatures of a quasilocalization, as we have shown in the main text for the confined Ising model. This effect 
is caused by the confining nature of the long-range interactions. 
Similarly to the case of the quantum Ising chain,
the frequency of oscillation $\nu$ is proportional to the external magnetic field strength $\nu\propto h$, and 
the dependence of the confinement length $\xi_{loc}$ on the parameters of the Hamiltonian~\eqref{Eq:XXZ_stag1}: 
as can be observed in Fig.~\ref{fig:XXZ_stag1}, we have $\xi_{loc}\propto {\epsilon}/{h}$, analogous to the Ising case upon the substitution $g\leftrightarrow \epsilon$. 

The system is extremely sensitive to the parameter $\epsilon$ that is proportional to the inverse mass of the quasi-particles. Indeed, as we increase $\epsilon$ toward the Heisenberg isotropic limit ($\epsilon=1$), the false vacuum becomes more unstable and the thermalization process is enhanced. 
It is natural to expect that in the planar phase ($|\epsilon|>1$), in which the excitations are massless, the effect of confinement disappears. 

The effect of the confinement can be also quantified investigating the entanglement entropy growth. In particular, from the density plot of the entanglement (insets in Fig.~\ref{fig:XXZ_stag1})  and from the average time-dependent entanglement entropy, defined as $S_{avg}=\sum_i S_i/L$ with $S_i$ the entanglement entropy computed across the boundary at site $i$, for several values of $h$ (panel in Fig.~\ref{fig:XXZ_stag2}), we can observe how the spreading of information in the system is slowed down by the presence of the confining interaction.
Even in this case, the oscillations of the kink in the middle can be viewed as Bloch-oscillations and studied analytically considering an effective Hamiltonian, obtained by projecting \eqref{Eq:XXZ_stag1} onto the one-kink (or two-kink) subspace.

\subsection{The lattice Schwinger model}

\label{sec_LSM}
In this Section we report some results which show that the effects of particle confinement on the string dynamics  of the lattice Schwinger model in Eq.~\eqref{eq:Schwinger} are similar to those of the linear potential between domain-walls in the quantum Ising chain with a tilted magnetic field. 

For this purpose let us consider ``string'' initial states, i.e., gauge-invariant eigenstates of the Hamiltonian in the non-interacting limit $w=0$, with a particle $q$ and an anti-particle $\bar{q}$ located at a distance $d$ along the chain, and let us measure the evolution of the electric flux spatial profile, {$\langle E_{j,j+1}(t) \rangle$}. 
Numerical simulations are performed with exact diagonalization techniques applied to the model obtained after integrating out the gauge field, i.e., a globally neutral system of fermionic charges with long-range Coulomb interactions \cite{Banks1976}.
\begin{figure}
    \centering
    \includegraphics[scale=0.25]{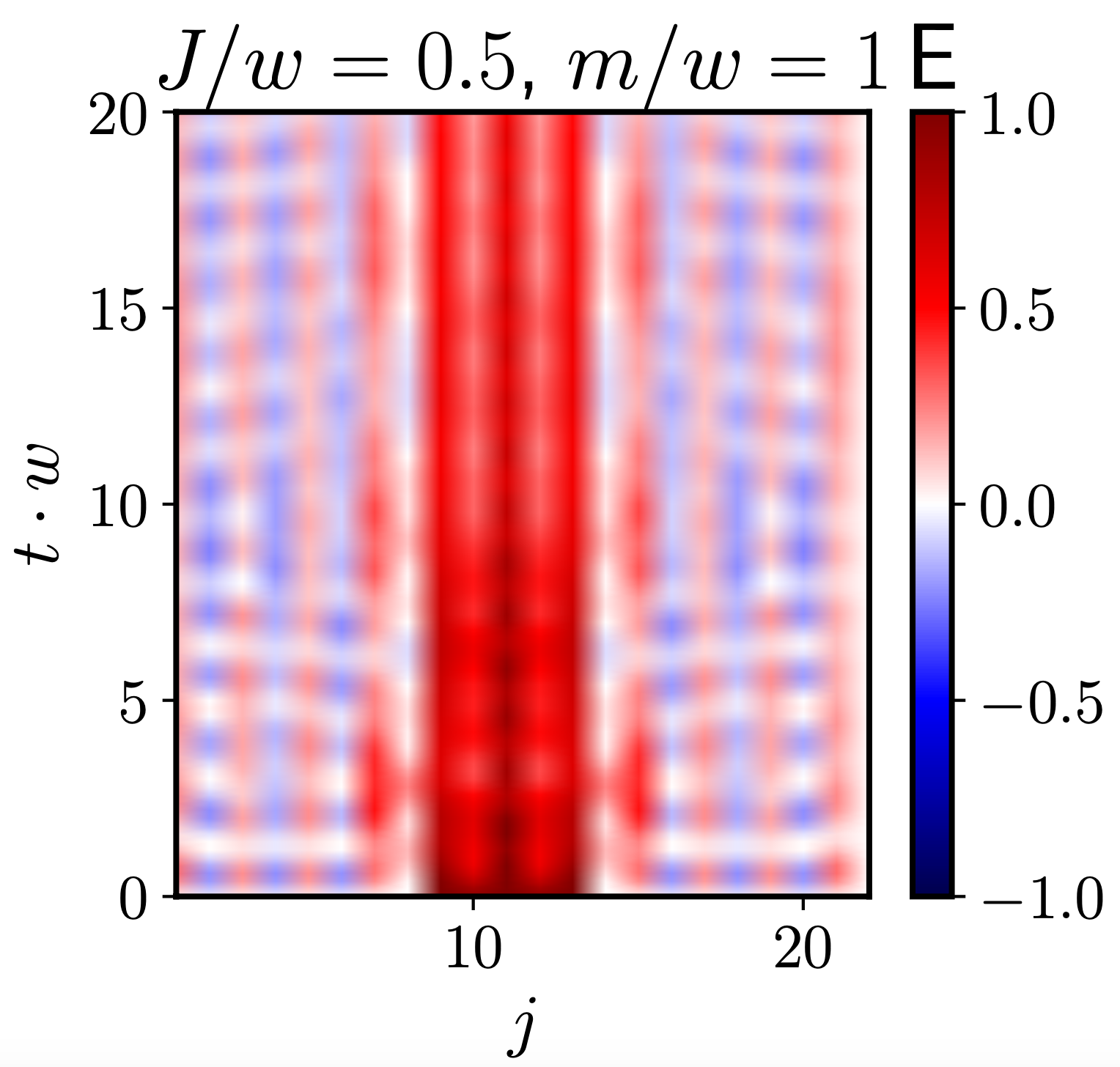}
    \caption{Space-time dependence of the electric field $\langle E_{j,j+1}(t) \rangle$ starting from a nonentangled initial state with one $q-\bar{q}$ pair, governed by the Hamiltonian~\eqref{eq:Schwinger} with the indicated values of the coupling $J/w$ and of the mass $m/w$.}
    \label{fig:string_Schwinger}
\end{figure}

Results are shown in Fig.~\ref{fig:string_Schwinger} for $d=5$. Away from the initial particles, vacuum fluctuations made up of virtual particle-antiparticle pairs appear, as signaled by the small coherent oscillations of the local electric field, cf. Refs. \onlinecite{KCTC,GorshkovConfinement,SuraceRydberg}. 
However, the spatial inhomogeneity of the electric field persists for long times, due to the suppression of string breaking, despite the sizeable strength $w=m$ of the interactions.

\bibliography{DWLR,EE,LGT,General}